\documentclass[12pt]{article}
\usepackage[english]{babel}
\usepackage[]{amssymb}
\newcommand{\be}{\begin{equation}}
\newcommand{\ee}{\end{equation}}
\newcommand{\ba}{\begin{eqnarray}}
\newcommand{\ea}{\end{eqnarray}}
\newcommand{\db}{\not{\hspace{-0.06 cm}}\partial}
\newcommand{\dsl}
  {\kern.06em\hbox{\raise.15ex\hbox{$/$}\kern-.56em\hbox{$\partial$}}}
  
\newcommand{\Dsl}{\not\!\! D}

\newcommand{\eeq}{\end{equation}}
\newcommand{\eeqarr}{\end{eqnarray}}
\begin{document}
\title{Vortices\footnote{Lectures given at the ``4th Chilean School
of Astrophysics, Cosmology and Gravitation" Valparaiso, October 2006.}}
\author{F.A.~Schaposnik\thanks{Associated with CICPBA}\\
{\normalsize\it Departamento de F\'\i sica, Universidad Nacional
de La Plata}\\
{\normalsize\it C.C. 67, 1900 La Plata, Argentina}}
\date{\hfill}
\maketitle
\abstract{I discuss in these lectures vortex-like classical solutions
to the equations of motion of gauge theories with spontaneous symmetry
breaking. Starting from the Nielsen-Olesen ansatz for the Abelian Higgs
model,  extensions to the case in which gauge dynamics
is governed by Yang-Mills and Chern-Simons actions are presented.
The case of semilocal vortices and also the coupling to axions
is analyzed. Finally, the connection between supersymmetry
and the existence of first order BPS equations in such models is
described.}

\newpage

\hfill{\begin{verse}
 { {\footnotesize \it Turning and turning in the widening gyre\\
  The falcon cannot hear the falconer}\\
 {\footnotesize   W.B.Yeats, The Second Coming\cite{Y}}}
\end{verse}}
These notes cover the topics of 4 lectures on vortex
solutions in spontaneously broken gauge theories and their
connection with supersymmetry, given at the {\it 4th Chilean School
of Astrophysics, Cosmology and Gravitation} held in Valparaiso last October.
The bibliography at the end of the notes is not complete: it only
refers to works containing results discussed in the lectures. A more complete list of
references can be found in some of the excellent books and reviews on this topic
\cite{Rev1}-\cite{Revn}.

\section{Nielsen Olesen vortices}
Vortices are very ubiquitous objets. Starting from the 1957 work
of Abrikosov \cite{Abrikosov} on superconductivity, belonging
to the area of condensed matter, they were then rediscovered  by
high energy physicists in a field theory context,
after the well-honnored paper by Nielsen
and Olesen \cite{NO}. In the present century    a revival took place due to
their cosmological interest in connection with cosmic superstrings
(see \cite{Polchinski} and references therein).

 Nielsen and Olesen \cite{NO} introduced
vortices in high-energy physics in an attempt to describe
classical string-like structures with the property of
having equations of motion identical with those of the
Nambu dual string.

In their work, Nielsen and Olesen started from a $3+1$ Lagrangian for an
Abelian $U(1)$ gauge theory coupled to a charged scalar with a
symmetry breaking potential and first found a classical
string-like (vortex) solution. Then they obtained the remarkable
result the action  is proportional to the area of the surface
swept out in space and time by such vortex. This result is the
basis of their identification of the Abelian Higgs Lagrangian with
the Nambu Lagrangian for the dual string \cite{nambu}.

By making a static axially
symmetric ansatz they reduced the Euler-Lagrange equations to a
pair of second order non-linear coupled differential equations in
the radial variable. Although they didn't find exact solutions (in
fact, analytical vortex solutions are not known) they gave arguments to
show that nontrivial solutions exist. They correspond to
vortex-like object carrying a magnetic field
in its core, where
the scalar field vanishes. The  magnetic flux is quantized (already
at the
classical level).

Such solutions look like type-II superconductivity vortices and
this is no coincidence: it is not difficult to see the connection
between the Abelian-Higgs model Lagrangian and the Ginsburg-Landau
fenomenological model for superconductivity, provided the Higgs
scalar in the relativistic field theory is identified with the
order parameter in the Ginzburg-Landau model. In order to have
type-II superconductivity the relation between the gauge coupling
constant $e^2$ and the symmetry breaking coupling constant
$\lambda$ should be  $e^2 < 8 \lambda$ (with the conventions
adopted below); for $ e^2 > 8 \lambda$ a complete Meisner effect
takes place (type-I superconductivity). The limiting point
 \be
 e^2 = 8 \lambda
 \label{bog}
 \ee
plays also an important role in the field theory/string theory context. In fact,
a surprising connection was
already notice in 1976, when it was observed \cite{dVS} that
relation (\ref{bog}) is precisely the same needed for the
supersymmetric extension of the model \cite{Fayet}-\cite{SalSt}.

The Abelian Higgs model dynamics in (3+1)-Minkowski space is
governed  by the action:
\begin{equation}
{  S}  = \int d^4x
\left(-\frac{1}{4}F_{\mu\nu}F^{\mu\nu} +
\frac{1}{2}(D_{\mu}\phi)^*(D^{\mu}\phi) -  V(|\phi|)
\right)
 \label{1}
\end{equation}
Here $F_{\mu\nu}$ is the field strength,
\begin{equation}
F_{\mu\nu} = \partial_\mu A_\nu - \partial_\nu A_\mu \; ,
\end{equation}
 $\phi$ is a complex scalar,
\begin{equation}
\phi = \phi^1 + i\phi^2 \; ,   \label{2}
\end{equation}
the covariant derivative is defined as
\begin{equation}
D_{\mu}= \partial_{\mu} + ieA_{\mu} \; ,  \label{3x}
\end{equation}
and the symmetry breaking potential taken in the form
\begin{equation}
V(|\phi|) =  \lambda \left(  |\phi|^2 - \phi_0^2 \right)
\end{equation}
Here $\phi_0$ is a real positive constant with [mass]  dimensions.

The resulting equations of motion are
\begin{eqnarray}
\partial^\mu F_{\mu\nu} &=& j_\nu \nonumber \\
D^\mu D_\mu \phi &=& -\frac{\delta V}{\delta \phi^*}
\end{eqnarray}
with the Higgs field current given by
\begin{equation}
j_\nu = \frac{ie}{2}(\phi \partial_\nu \phi^* - \phi^* \partial_\nu \phi) + e^2 |\phi|^2
A_\nu
\label{jota}
\end{equation}

The Nielsen-Olesen strategy
to construct nontrivial regular solutions to the
equations of motion with finite energy (per unit length)
starts from a trivial (constant) solution
and implies the following steps:

~

\begin{enumerate}
\item{\underline{Trivial}}  zero-energy solution
$
|\phi |= \phi_0  \; , \;\;\; A_i = {0}
$

~

\item {\underline{non-trivial} but  \underline{singular}
 solution}  (fluxon with $N$ units of magnetic flux) which in polar coordinates
 reads
\[
\phi =  \phi_0 \exp(i N\varphi) \; , \;\;\; A_i =
N \partial_i \varphi\; , \;\;\;
\varepsilon_{ij}F_{ij} = 4\pi {N\delta^{(2)}(\vec x)}
\]
\be
\frac{1}{4\pi} \oint \varepsilon_{ij}F_{ij} = N \in \mathbb{Z}
\label{top}
\ee
This configuration is a gauge transformed of the trivial one.
The appropriate
gauge group element to consider is
\be
g_N(\varphi) = \exp(iN\varphi)
\label{gn}
\ee
 which is ill-defined at the origin.

\item{ \underline{Regular}  Nielsen-Olesen vortex ansatz}
\be
\phi =  f(r) g_N(\varphi) { }  \; , \;\;\; A_i =
   {a(r)} \frac{1}{i}\, g_N^{-1} (\varphi)\partial_i g_N(\varphi) \\
 \label{reg}
\ee
with
\be
f(0)=a(0)= 0 \; , \;\;\; f(\infty) = \phi_0  \; , \;\;a(\infty) = 1
\label{cond}
\ee

\end{enumerate}
This ansatz leads to the same magnetic flux as that in (\ref{top}) but
there is no singularity  in view of the
boundary conditions that one imposes at the origin on $f$ and $a$.
Conditions at infinity guarantee that each term
in the Lagrangian goes to zero at infinity.

Being axially symmetric, ansatz (\ref{reg}) can be interpreted as a magnetic tube
with quantized flux (a vortex). Being  $z$-independent, such a solution has infinite energy.
However, if we define an energy $E$ per unit length, conditions (\ref{cond}) ensure its
finiteness,
\be
E =
\int d^2x \left( \frac{1}{4}F^2_{ij} + \frac{1}{2}\vert D_i\phi\vert^2
+ \lambda(\vert\phi\vert^2 - {\phi_0}^2)^2 \right)
\ee
Being $z$ independent, such a static vortex solution can be also interpreted as
a soliton solution in $2+1$ dimensions. Because it depends only on $x,y$, it can also be
thought as an instanton solution in d=2 Euclidean dimensions. In any two of this cases,
space at infinity can be seen as a circle $S^1_\infty$. But also the gauge group, $U(1)$,
can be identified with a circle $S^1$ so that one can see $g_N$ as establishing a map
\be
g_N: S^1_\infty \to S^1
\ee
Such maps are classified into homotopy classes $\Pi_1$ (where the subindex refers to
the dimension $n$ of the sphere at infinity, in this case $n= 1$). Now, one has
\be
\Pi_1(S^1) = \mathbb{Z}
\ee
and in this way one connects the vortex magnetic
flux (\ref{top})with a topological charge.

Unlike monopoles or instantons,  no analytic solution to the
equations of motion is known. However, numerical solutions can be
easily constructed and there are also proofs that solutions exist
(see \cite{JT} and references therein).
The asymptotic behavior of the vortex solutions is
\ba
a(r) &=& -\frac{1}{e} + mr K_1(-m_V r)\nonumber\\
\phi(r) &=& \phi_0 + O\! \left(\exp(-m_\phi r)\right)
\ea
where $m_V$ and $m_\phi$ are the masses of the vector and scalar
particles respectively,
\ba
m_V &=& e\phi_0 \label{ma}\\
m_\phi &=& 2 \sqrt{2\lambda }\phi_0 \label{mi}
\ea
 Note that
when condition (\ref{bog}) is satisfied both masses coincide, this being
the first indication that the model can be viewed as describing
the bosonic sector of an $N=2$ supersymmetric  theory, with the scalar and the gauge
field in the same supermultiplet. In fact when $m_V \sim m_\phi$ we have a well defined
vortex line or a well defined string of extension $1/m_V \sim 1/m_\phi$. Inside
the string one basically has magnetic field while the scalar field vanishes. Outside
the string it is the magnetic field which is nearly zero while the Higgs scalar is
practically in
its vacuum value $|\phi| = \phi_0$.

Explicit solutions in the whole $r$ range can be more easily found observing that,
under magic condition (\ref{bog}) the second order
Euler-Lagrange equations can be reduced to first-order ones. There are different ways
of arriving to this result. One was discovered in \cite{dVS} and starts by
demanding that the space-space energy-momentum components ($T_{rr}$ and
 $T_{\varphi\varphi}$) vanish. The other one establishes a bound (Bogomol'nyi bound)
 for the vortex energy
 (per unit length) which is saturated whenever (\ref{bog}) holds
 \cite{Bog}. Being configurations static, such a bound also corresponds to a bound for the
 action and hence they satisfy the equations of motion.

\subsection*{The Bogomol'nyi bound}

Let us see how Bogomol'nyi found in ref.\cite{Bog} a bound for the energy starting from formula
(\ref{bog}). To this end, let us write the scalar field in terms of its real and
imaginary components
\be
\phi =  (\phi^1,\phi^2) = (\phi^a) \; , \;\;\; a=1,2
\ee
and scale fields and coordinates according to
\be
\phi^a \to \phi_0 \phi^a \; ,  \;\;\; A_i \to \phi_0 A_i \; , \;\;\; x \to
\frac{1}{e\phi_0} y
\ee
so that the energy reads now
\be
E = \phi_0^2 \int d^2 y \left(\frac{1}{4}F^2_{ij} + \frac{1}{2} D_i\phi^a
D_i\phi^a
+ \frac{\lambda}{e^2}(\vert\phi\vert^2 - 1)^2
\right)
\ee
where the covariant derivative $D_i^{ab}$   is defined as
\be
D_i^{ab}  = \delta^{ab} \partial_i   + \varepsilon^{ab} A_i
\ee

One easily sees that the energy can be rewritten as

\newpage
\ba
E &=& \phi_0^2 \int d^2 y\left(
\frac1 4 \left( F_{ij} \pm \frac{1}{2} \varepsilon_{ij}
 (\phi^a\phi^a -1)^2\right)^2 \right.
 \nonumber\\
 &&+
 \frac{1}{4}
 \left( D_i \phi^a \mp \varepsilon^{ab}\varepsilon_{ij} D_j\phi^b
  \right)^2  + \left(\frac{\lambda}{e^2} -
 \frac{1}{8} \right)(\phi^a\phi^a - 1)^2
 \nonumber\\
 & &
 \left. \pm \frac{1}{4} \left(\varepsilon_{ij}F_{ij}
 \mp \frac{1}{2}\varepsilon^{ab}\varepsilon_{ij} \partial_i(\phi^a D_j\phi^b)
 \right)
\right)
\ea
The term
\be
\frac{1}{2}\varepsilon^{ab}\varepsilon_{ij} \partial_i(\phi^a D_j\phi^b)
\ee
gives no contribution to the energy, as can be seen by using the Gauss-Stokes theorem
and the fact that $D_i\phi^b = 0$ on $S^1_\infty$. Then, using (\ref{top}) we have
\ba
E \!\!\!&&= \phi_0^2 \int d^2 y\left(
\frac1 4 \left( F_{ij} \pm \frac{1}{2} \varepsilon_{ij}
 (\phi^a\phi^a -1)^2\right)^2
 \right.\nonumber\\
 & & \left.+ \frac{1}{4} \left( D_i \phi^a \mp
 \varepsilon^{ab}\varepsilon_{ij}
 D_j\phi^b \right)^2 + \left(\frac{\lambda}{e^2} -
 \frac{1}{8} \right)(\phi^a\phi^a - 1)^2
\right)
\pm \phi^2_0 \pi N \nonumber \\
\label{bound}
\ea
Were the third term absent, the energy would just be the sum of to positive
definite terms plus a term proportional to the topological charge. This is precisely
what happens for the Bogomol'nyi point (\ref{bog}) for which one then has
\be
E \geq \phi_0^2 \pi |N|
\label{boundI}
\ee
The bound is attained when the two perfect squares vanish, and this leads to the so-called
Bogomol'nyi
or BPS equations for vortices,
\ba
F_{ij} &=& \mp \varepsilon_{ij} \left( \phi^2 -1\right) \nonumber\\
D_i\phi^a &=& \pm \varepsilon^{ab}\varepsilon_{ij} D_j\phi^b
\label{devs}
\ea

The ''PS'' refers to M.~K.~Prasad and C.~M.~Sommerfield who found
exact monopole solutions for the non-Abelian version of the model
for a particular (Bogomol'nyi) value of the coupling constant
\cite{PS} ($\lambda=0$). That their solution corresponds to a
Bogomol'nyi bound was first noted in \cite{ColN}.

As mentioned above, eqs.(\ref{devs}) were independently obtained
 by asking the
stress tensor components $T_{ij}$ to vanish. This is consistent
 with the eqs.
of motion only if $\lambda = 8e^2$. When matter interaction
exceeds that of the electromagnetic one ($\lambda > 8e^2$) vortices
attract whereas in the opposite case they repel. In the former case
a system of many superimposed vortices  decays into separated vortices.
In the case $\lambda = 8e^2$ vortices do not interact and
this is consistent with thew fact that the energy is proportional to the
topological charge. An exact (numerical) solution to the system
(\ref{devs}) was presented in \cite{dVS}. The existence theorem for vortex (and monopole) solutions can be found
in   A.~Jaffe and C.~H.~Taubes book on Vortices and Monopoles \cite{JT}.

\section*{Chern-Simons vortices}
As explained above, if one defines the Maxwell-Higgs system in
$d=2+1$ dimensions, vortices can be seen as static finite energy
(solitonic) solutions of the equations of motion. Now, as it is
well-known, there is in three dimensional space time a second
Lorentz invariant term which can be added to the Maxwell term,
namely the Chern-Simons term which, for a $U(1)$ gauge theory
reads
\be L_{CS} = \frac{\mu}{4} \varepsilon^{\nu\alpha\beta} F_{ \nu\alpha}
A_\beta
\ee
Here $\mu$ is a parameter with mass dimensions usually called  the
topological mass. This name is due to the following important
fact related to the Chern-Simons term \cite{DT}-\cite{DJT}.
Consider the U(1) gauge theory with Lagrangian
\be
L_{M+CS} = -\frac{1}{4} F_{\mu\nu}F^{\mu\nu} +
 \frac{\mu}{4} \varepsilon^{\nu\alpha\beta} F_{ \nu\alpha}
A_\beta
\ee
Now, from the eqs. of motion for this Lagrangian,
\be
\partial_\mu F^{\mu\nu} + \mu {^*}\!{F^\nu} = 0
\ee
with the dual field strength ${^*}\!{F^\nu}$ given by
\be
{^*}\!{F^\nu} = \frac 1 2 \varepsilon^{\nu\alpha\beta}F_{\alpha\beta}
\ee
one can find
\be
\Box {^*}\!{F^\nu}  + \mu {^*}\!{F^\nu} = 0
\ee
showing that the gauge field is massive.

Clearly the field equations are gauge invariant but the Lagrangian is not. Under
a change
\be
A_\mu \to A_\mu + \partial_\mu \Lambda
\ee
the Chern-Simons Lagrangian changes by a total derivative
\be
L_{CS} \to L_{CS} + \frac{\mu}{2} \partial_\alpha \left( {^*}\!{F^\alpha} \Lambda
\right)
\ee
The action is however gauge-invariant since the field strength fall
off at large distances and one accepts appropriate $\Lambda'{\rm s}$.

Let us then consider the Maxwell-Higgs model with the addition of a
CS term, with Lagrangian
\begin{equation}
{  S}  = \int d^3x
\left(-\frac{1}{4}F_{\mu\nu}F^{\mu\nu} +
\frac{\mu}{4} \varepsilon^{\nu\alpha\beta} F_{ \nu\alpha}
A_\beta
+ \frac{1}{2}(D_{\mu}\phi)^*(D^{\mu}\phi) -  V(|\phi|)
\right)
 \label{1CS}
\end{equation}
The equations of motion now read
\begin{eqnarray}
\partial^\mu F_{\mu\nu} + \mu {^*}\!{F_\nu} &=& j_\nu \label{arriba} \\
D^\mu D_\nu \phi &=& \frac{\delta V}{\delta \phi^*} \label{abajo}
\end{eqnarray}
Now, integrating the temporal component  of eqs.(\ref{arriba}) (the  Gauss law)
in the plane
one has
\be
\int d^2x \left(\partial^i F_{i0} + \mu \varepsilon_{ij}F^{ij}
\right) =  \int d^2x j_0
\ee
The first term in the l.h.s. is a surface term and gives no contribution. The second
one is just the magnetic flux and that in the r.h.s. is identified
with the electric charge of the configuration. One then has
\be
\mu \Phi = Q
\label{rela}
\ee
with
\ba
\Phi = \frac{1}{2}\int d^2x \varepsilon_{ij}F^{ij} \; , \;\;\; Q = \int d^2x j_0
\ea
If one looks for static configurations, the current component $j_0$ is just
\be
j_0 =  e^2 |\phi|^2
A_0
\ee
We then see that either $A_0 \ne 0$ or we cannot have vortex-like solutions like
those discussed for the pure Maxwell-Higgs system. Interestingly enough, if vortex
solutions exist,  magnetic flux and electric charge are related according to
(\ref{rela}). Being the magnetic flux related to the topological charge, the electric
charge
is then also quantized already at the classical level.

In fact, it is very easy to extend the ansatz (\ref{reg}) in order to include
an $A_0$  \cite{PK},
\be
\phi = f(r) \exp(in\varphi) \; , \;\;\; A_i = na(r) \partial_i \varphi
\; , \;\;\; A_0 = a_0(r)
\label{asero}
\ee
Apart from conditions (\ref{cond}), one should ask
\be
a_0(0) = a_0(\infty) = 0
\ee
As it is to be expected in view of our previous discussion,
the vector mass is modified due to the presence of the Chern-Simons term
(while the scalar field mass remains unchanged)
\ba
m_{A\pm} &=& \sqrt{ \frac{\mu^2}{4}  + e^2\phi_0^2} \pm \frac{\mu}{2} \label{ma2}\\
m_\phi &=& 2 \sqrt{2\lambda }\phi_0 \label{mi2}
\ea
Using the axially symmetric ansatz, numerical solutions can be found. The
behavior of the Higgs scalar
and the spacial components of the gauge field is qualitatively the same as in the
pure Maxwell-Higgs case. Concerning the electric field, it vanishes at the
origin and has its maximum value at finite $r$. The energy $E$ of the vortex can be seen
to be $E_\pm \sim m_{A\pm}^2 \log\left(e^2/m_{A\pm}^2 \right)$ and hence the $m_{A-}$
solution has lower energy.

No Bogomol'nyi bound for the energy can be constructed when the CS
term is included. Hence, there are in this case (and with this
field content) no BPS first order equations for the
Maxwell-Higgs-Chern-Simons model.

There is however a way out to this problem; one is to modify the
field content. The other one has been discovered in two
simultaneous works \cite{Hong}-\cite{JW}. Basically, it consist in
not including the Maxwell term -that is, the gauge field dynamics
is solely  governed   just the Chern-Simons action- and changing
the symmetry breaking potential from the usual fourth order one to
a sixth order potential which is acceptable from the point of view
of renormalizability since the model is defined in $2+1$
space-time. The model can be seen as the truncation at large
distances and low energies where the lower derivative Chern-Simons
term dominates the higher derivative Maxwell term.

The Lagrangian of the model is
\begin{equation}
{  S}  = \int d^3x \left(
 \frac{\mu}{4}
\varepsilon^{\nu\alpha\beta} F_{ \nu\alpha} A_\beta +
\frac{1}{2}(D_{\mu}\phi)^*(D^{\mu}\phi) -  V^{(6)}(|\phi|) \right)
 \label{2CS}
\end{equation}
Although one can include in the potential an arbitrary sum of
fourth-order and sixth order potential, Bogomol'nyi equations will
arise for a particular combination that can be written in the form
\be
V^{(6)} = \frac{e^4}{8 \mu^2} |\phi|^2\left(|\phi|^2 - \phi_0^2
\right)^2
\ee
There is no independent coupling constant in this potential because it
has been already defined at the Bogomol'nyi point. Note that being defined in a $d=3$ dimensional
theory, it corresponds to a renormalizable potential.

Variation of the action yield to the field equations
\ba
\frac{\mu}{2}  \varepsilon^{\alpha\beta\gamma}F_{\beta\gamma} &=& j^\alpha
\label{gau}\\
D^\mu D_\mu \phi &=& -\frac{\delta V^{(6)}}{\delta \phi^*}
\label{giu}
\ea
where $j^\alpha$ is still given by (\ref{jota}). The time component of
eq.(\ref{gau}) is the Gauss law,
\be
 \mu  F_{12} = j_0
\ee

Noting that there is no metric in the Chern-Simons action
(in this sense it is a
topological Scwartz like action \cite{BBRT})
 one concludes that
it does not contribute to the energy momentum tensor,
which is defined as
\be
T_{\mu\nu} = 2\frac{\delta S}{\delta g^{\mu\nu}}
\ee
Thus, the energy functional for static field configurations reads
\be
E = \int d^2x  \left(
\frac{1}{2}|D_i\phi|^2 + \frac{e^2}{2} A_0^2 |\phi|^2 + V(|\phi|^2)
\right)
\label{energy}
\ee
In the static case $j_0 = e^2 |\phi|^2
A_0$ so that we can use the Gauss law to solve for $A_0$ giving
\be
A_0 = \frac{\mu}{2e^2} \frac{F_{12}}{|\phi|^2}
\ee
Inserting this in (\ref{energy}) one can easily accommodate the energy \`a la Bogomol'nyi,
  a sum of squares
plus the a term proportional to the magnetic flux,
\ba
E &=& \int d^2x \left(
\left|(D_1 \pm i D_2)\phi\right|^2   + \left| \frac{\mu}{2e} \phi^{-1} F_{12}
 \pm \frac{e^2}{\mu}\phi^*(\phi_0^2 - |\phi|^2 \right|^2
\right) \nonumber\\
&&
\mp \frac{e\phi_0^2}{2} \Phi
\label{boundi}
\ea
where we have discarded a surface term which can be converted to a line integral
vanishing for finite energy configurations.
Then, for a fixed value of the flux, $\Phi = 2\pi N/e$  we have a bound for the energy,
\be
E \geq \pi e\phi_0^2 |N|
\label{baundii}
\ee
The bound is saturated when the following first-order equations hold,
\ba
D_1 \phi &=&  \mp i D_2\phi \nonumber\\
e F_{12} &=& \mp \frac{m^2}{2}  \frac{|\phi|^2}{\phi_0^2}
\left( 1 -\frac{|\phi|^2}{\phi_0^2} \right)
\ea
where $m = 2e^2\phi_0^2/|\mu|$. This mass coincides with the vector and the scalar field
masses $m = m_A= m_\phi$, this again signaling some relation with $N=2$ supersymmetry.

Being a minimum for the energy, a solution of these equations
 also solve are
also static solutions of the
eqs. of motion.

Axially symmetric solutions have been studied
using an ansatz of the form (\ref{asero}).
Concerning the Higgs, the field profile is qualitatively analogous
tho those for the Maxwell-Higgs system. As for the magnetic and electric fields,
as it happens for the Maxwell-CS-Higgs model, they are concentrated in a ring
with its maximum occurring when  $|\phi|^2 = \phi_0^2$.


\section*{Semilocal vortices}
Suppose that one replaces the complex scalar field in the
Maxwell-Higgs model by an $SU(2)$ doublet  $\phi =
(\phi^I,\phi^{II})$ \cite{VA}. The action is
\begin{eqnarray}
{  S} \!\!\!\!&=&  \!\!\!\!  \int d^4x
\left(-\frac{1}{4}F_{\mu\nu}F^{\mu\nu} +
\frac{1}{2}(D_{\mu}\phi^I)^*(D^{\mu}\phi^I)
+
\frac{1}{2}(D_{\mu}\phi^{II})^*(D^{\mu}\phi^{II})- \right.\nonumber\\
&&
 \left.  V\left(|\phi^I|^2 + |\phi^{II}|^2 \right)
 \vphantom{\frac{1}{4}}
\right)
 \label{1SL}
\end{eqnarray}
This action is not only invariant under a local $U(1)_{Local}$
gauge group but also under
a global $SU(2)_{Global}$ group. The total symmetry group is then
\be
G = SU(2)_{global} \times U(1)_{local}
\ee
The potential has an $O(4)$ symmetry and when it corresponds to a symmetry breaking one
the vacuum manifold is a three sphere defined by
\be
{\phi^I_1}^2 +  {\phi^I_2}^2 + {\phi^{II}_1}^2 +  {\phi^{II}_2}^2 = \phi_0^2
\ee
which is simple connected so that there are no topological string solutions
($\Pi_1(S^3) = 0$). Now
$SU(2)$ global rotations in the direction of the constant doublet
$(\phi_0^I,\phi_0^{II})$
leave invariant the vacuum so that  the symmetry breaks to a global $U(1)_G$,
\be
SU(2)_{global} \times U(1)_{local} \to U(1)_{global}
\ee

Flux-tube solutions to the equations of motion of such a model have
been presented in \cite{VA} where, in view of the
symmetry scheme,  they have been called ``semilocal strings''. In order to
understand the nature of such solutions, let us note that the equations of motion
for the model are formally the same as those for the Abelian Higgs model with just
one scalar, except that now such scalar has been replaced by an SU(2) doublet.
Then any (Nielsen-Olesen) vortex solution
\be
\phi = f(r)\exp(iN\varphi)\; ,  \;\;\; A_i = a(r)\partial_i\phi
\ee
of the ordinary model can
be extended to one of the present
one by introducing an $SU(2)$ doublet $\Phi_0$ of unit norm,
\be
\Phi_0 =
\left (\matrix{ a \exp(i\varphi^I)\cr
\sqrt{1 - a^2} \exp(i\varphi^{II})\cr }\right)
\ee
and writing
the semilocal solution in the form
\be
\Phi = f(r) \Phi_0  \; , \;\;\; A_i = a(r)\partial_i\phi
\label{solmul}
\ee
Here  $0 \leq  a  \leq 1$.

Now, in order to have finite energy, covariant derivatives of $\Phi^I$ and
$\Phi^{II}$ should vanish at infinity, this implying
\ba
\lim_{r \to \infty} D_i\phi^I = 0
 && \rightarrow (i\partial_i \varphi^I -iN \partial_i \varphi) = 0\nonumber\\
\lim_{r \to \infty} D_i\phi^{II} = 0
 && \rightarrow (i\partial_i \varphi^{II} -iN \partial_i \varphi) = 0
\ea
so that the two phases can differ just in one constant,
\ba
\varphi^I &=& N\varphi + C\nonumber\\
\varphi^{II} &=& N\varphi
\ea
and we have that asymptotically, finite energy configurations should take
the form
\be
\Phi(\infty) = \phi_0\exp(iN\varphi)
\left (\matrix{ a \exp(iC)\cr
\sqrt{1 - a^2}  \cr }\right)
\ee
Although there are still a $U(1)_{Global}$ rotations of this configuration that
pass to another one which is still a minimum of the potential, the energy
of the transformed configuration should be infinity because of the change in
the covariant derivatives. That is, the space that label finite energy configurations
does not correspond to the vacuum manifold ${\cal V}$ (the zeroes of the potential) but
to the gauge orbit from any point $\Phi_0 \in {\cal V}$. In this sense the relevant
homotopy group is $\Pi_1(U(1)) = \mathbb{Z}$. Thus configurations with different
winding numbers are separated by infinite energy barriers, an information that
is not contained solely in the vacuum manifold $ {\cal V}$ but also in the kinetic
energy term. Now, since $\Pi( {\cal V}) = 0 $ it is not topologically guaranteed
that a configuration characterized by some number $N$ is stable.

In summary, although the vacuum manifold is simply connected and a field configuration
that winds at infinity can be unwind without any cost of potential energy, this does
not guarantee that there is no cost of the gradient part of the energy and in fact, this
is what happens due to the non-triviality of $\Pi(G_{Local}/{H_{Local}})
= \Pi_1(U(1)) = \mathbb{Z} $.

Stability of the configurations have been studied in \cite{Hind}-\cite{AV2}. Again,
the Bogomol'nyi point plays a central role;for $e^2 = 8\lambda$ one can repeat the
Bogomol'nyi argument and write a bound for the energy, which is just that given in
eqs.(\ref{bound})-(\ref{boundI}) but where the single complex scalar is
 replaced by an SU(2) doublet. Being a local minimum, the solution
 written in (\ref{solmul}) with $f(r)$ and $a(r)$ the radial functions for
the Nielsen-Olesen solution is classically stable. One should mention the existence
of zero-modes \cite{H}  which will be described below.

\section*{Non Abelian vortices}

The first reference to (local) vortex solutions in non-Abelian gauge theories can be
found in the Appendix of Nielsen and Olesen pioneer paper \cite{NO}. However,
it was rapidly recognized that the boundary conditions imposed in \cite{NO}
have solutions that in some cases could be deformed continuously to topologically
trivial ones \cite{Man}.
It was then  understood that in order to have topologically stable solutions,
symmetry breaking has to be complete and hence more than one
Higgs scalar has to be introduced. This fact was discussed in detail in
\cite{deVe}-\cite{Schw} where it was observed that one needs $\Phi_1(G_L/H)$  to
be nontrivial. For example, for $SU(2)$ gauge theory, just one Higgs field in the
adjoint representation is not enough: one needs a second one and a choice
of symmetry breaking potential so that the only residual symmetry group element
is just the unit matrix in the adjoint so that $H= \mathbb{Z}_2$ and
$\Phi_1(SU(2)/\mathbb{Z}_2) = \mathbb{Z}_2$ ($\mathbb{Z}_N$ when the gauge group is
$SU(N)$).

Let us discuss in some detail the necessary conditions on the vacuum manifold in order
to have stable  vortices when the  gauge group is $G=SU(N)$. To begin with,
the election of a representation for
the Higgs field is crucial. Indeed, if we choose the fundamental representation of
$SU(N)$ or any other
 faithful representation (i.e., isomorphic to the group) the fundamental homotopy
 group $\Pi_1(G/H) = \Pi_1(SU(N)/I) = 0 = \Pi_1(SU(N)/I)$ since $SU(N)$
 is simply connected. The situation is different if the Higgs fields are in the adjoint,
 which does not correspond to a faithful representation of $SU(N)$.

 Noting that elements $\Omega_m^0$ such that
 \be
 \Omega_m^0 = \exp(2\pi i m/N) I \;, \;\;\; m=0,1, \ldots  N-1 \nonumber\\
\ee
can be taken as the elements of the Abelian group $\mathbb{Z}_N$,
one can define representatives of each homotopy class in the form
\be
\Omega_m(\varphi + 2\pi) = \exp(2\pi i m/N) \Omega_m(\varphi)
\label{omega}
 \ee
In brief, although we say that the symmetry is
 completely broken, there is the residual symmetry $H = \mathbb{Z}_N$ and then
 \be
\Pi_1(G/H) = \Pi_1(SU(N)/ \mathbb{Z}_N ) =  \mathbb{Z}_N
\ee
It should be noticed that, taken as a gauge transformation
group element, $\Omega_m(\varphi)$ is singular,
are the non-Abelian counterpart of
$\exp (im\varphi)$, for (Abelian) Nielsen-Olesen vortices.

Let us now discuss the number of Higgs scalars (in the adjoint) that one
has to introduce to break the symmetry so as to have topologically non
trivial solutions. We start by
recalling  the standard Cartan-Weyl basis,
\begin{eqnarray}
[H_i,H_j] &=& 0 \; , \;\;\; j=1,2,\ldots , N-1
\nonumber\\
~
[ H_i, E_{\alpha} ] & = &  r_{i\alpha}
E_\alpha \nonumber\\
~[E_\alpha,E_\alpha] &=& r_{i\alpha} H_i \nonumber\\
~[E_\alpha,E_\beta] &=&
\left\{\matrix{ N_{\alpha\beta} E_\gamma ~{\rm if}~  r_\gamma= r_\alpha +
r_\beta \cr {\rm otherwise} \cr
}
\right.
\ea
Here $r_\alpha$ is a root vector associated with $E_\alpha$,
$r_{i\alpha}$ one of its components.

It has been proved  that in order to completely break the symmetry and
have topologically non-trivial solution it is enough
to have one
Higgs field $\Psi$ in the Cartan subalgebra (the one generated by the $H_i$'s)
and $N-1$ fields $\Phi^A$ in the orthogonal complement
\cite{deVegS1}-\cite{deVegS2}.
With this in mind, let us take as the gauge Higgs action
\be
S = {\rm Tr}\!\! \int \!\! d^4x\!\!
\left( \!\!-\frac{1}{2} F_{\mu\nu}F^{\mu\nu} +
\sum_{A=1}^{N-1} D_{\mu}\Phi^A D^{\mu}\Phi^A
+ D^\mu\Psi D_\mu\Psi - V(\Phi,\Psi)
\right)
 \label{1na}
\end{equation}
where
\ba
A_\mu &=& A_\mu^a t^a \; , \;\;\; D_\mu = \partial_\mu + e[A_\mu,~] \nonumber\\
F_{\mu\nu} &=& \partial_\mu A_\nu - \partial_\nu A_\mu + e[A_\mu,A_\nu] \nonumber\\
~ [t^a,t^b] &=& i f^{abc} t^c \; , \;\;\; {\rm Tr} t^a t^b = \frac{1}{2} \delta^{ab} \; , \;\;\; a,b= 1,2,\ldots , N^2-1
\ea
The potential should be chosen so as to ensure complete symmetry breaking.

 A nonabelian extension of the
Nielsen-Olesen ansatz can be then proposed in the form
\begin{eqnarray}
\Phi^A &=& f^a (r)
\Omega_n^{-1}(\varphi)(E_{a_A} + E_{a_A}) \Omega_n(\varphi) \; , \;\;\; A=1,2,\ldots, N-1
\nonumber\\
\Psi &=& \sum_{j=1}^{N-1}C_j H_j\nonumber\\
A_\varphi  &=& \frac{n}{e}\, a(r) \mathbf{M} \; , \;\;\; A_r = 0 \nonumber\\
\mathbf{M} &=&  {\rm diag} (1/N,1/N, \ldots, 1/N, (1-N)/N)
\label{super}
\end{eqnarray}
where $\mathbf{M}$ belongs to the Cartan subalgebra. Note that $\Psi$ is taken as a constant
everywhere.
One can see that with this ansatz all fields $\Phi^A$ satisfy the same
equation of motion which, as it happens with the gauge field one, coincides
with those for the Abelian Higgs model so that their properties are the same
as those already discussed  for   Nielsen-Olesen vortices.

An ``electromagnetic'' tensor ${\cal F}_{\mu\nu}$ can be introduced in order
to characterize
the vortices. Following \cite{deVegS1}-\cite{deVegS2} we choose
\be
{\cal F}_{\mu\nu} = \frac{{\rm Tr}
(\mathbf{M}  F_{\mu\nu})}{{\rm Tr} (\mathbf{M}^2)}
\label{ele}
\ee
Then, the  flux associated to the magnetic field ${\cal F}_{12}$ reads, for
ansatz (\ref{super}),
\be
\Theta = \frac{1}{2} \int d^2x  \varepsilon_{ij}{\cal F}_{ij}=
 \frac{1}{\sqrt Ne}2\pi n
\label{flujo}
\ee

It is important to stress that ansatz (\ref{super}) in fact works for any $n = 2k
+ m$, with  $m=1,2,\ldots, N-1$ and $k\in\mathbb{Z}$. What is happening is that
the magnetic flux is not a topological number as
in the Abelian case. Indeed, as explained above,  the topological charge is
an element of $\mathbb{Z}_N$ label in this case by $m$.

\subsection*{The Nonabelian Bogomol'nyi bound}

A Bogomol'nyi bound for the nonabelian vortices discussed above was found in  \cite{CLS}.
We here sketch the $SU(2)$ case which can be easily extended to $SU(N)$. As explained
above, the potential should be chosen so as to completely break the symmetry. One can
  take
a potential of the form
\be
V(\Phi,\Psi) = \frac{1}{4}{\rm Tr}\left(  {g}   (\Phi^2 - \Phi_0^2)^2 +
 {g'}  (\Psi^2 - \Psi_0^2)^2 +  {g''}  (\Phi \Psi)^2
\right)
\ee
 minimum for the energy will take place when
\be
 {\rm Tr} (\Phi \Psi) = 0
\ee
The Higgs field  $\Psi$ in  (\ref{super}) does not  play
any dynamical role since
the ansatz corresponds to the condition
\be
D_i\Psi = 0
\label{nula}
\ee
everywhere. Then, the energy $E$ (per unit length) of a general configuration
is always bounded by that satisfying the ansatz (\ref{super})  according to
\ba
E \geq {\Phi_0^2}  {\rm Tr } \int d^2x \left(
 \frac{1}{2}F^2_{ij} +   D_i\Phi D_i\Phi +
   + \frac{g}{4e^2} \lambda(\Phi^2 - 1)^2
\right)
\label{pc}
\ea
Now, as usual, the r.h.s. in (\ref{pc}) can be written as a sum of perfect squares
plus additional surface terms/. After some work one finds
\ba
E & \geq &{\Phi_0^2}    \int d^2x \left(
\left(\frac{1}{4}{\vec F}_{ij}  + \frac{\alpha}{1 + \alpha^2} \varepsilon_{ij}(\Phi^2 - 1)
 \vec \Psi \right)^2 +
\right.
\nonumber\\
&&  \frac{1}{2(1 + \alpha^2)}\left( D_i\vec \Phi + b \varepsilon_{ij}
\vec \Psi \wedge  \vec{D_j\Phi}
 \right)^2 - \frac{1}{2(1 + \alpha^2)}  \varepsilon_{ij} \vec F_{ij}.\vec \Psi
+ \nonumber\\
&&  \left.
 \left (\lambda - \frac{\alpha^2}{2(1 + \alpha^2)}
 \right)(\Phi^2 - 1)^2
 \right)
 \label{81}
\ea
with $\alpha$, an arbitrary constant, should be chosen so that the term in
the last line is semipositive definite. This means
\ba
  |\alpha| &=& 1 \;\;\; {\rm if} \;\;\; \lambda \geq \frac{1}{8}
\nonumber\\
\frac{\alpha^2}{2(1 + \alpha^2)} &=& \lambda
\;\;\; {\rm if} \;\;\; \lambda \leq \frac{1}{8}
\ea
Being the last term in the second line of (\ref{81}) related to the magnetic flux
(\ref{flujo})
  the energy   bounded can be written as
\ba
E &\geq& \mp \frac{\Phi_0^2}{2} \pi n
   \;\;\;     {\rm if} \;\;\;  \lambda \geq \frac{1}{8}
\nonumber\\
E &\geq& \mp  \sqrt{2\lambda} \Phi_0^2 2\pi n
 \;\;\; {\rm if} \;\;\; \lambda \leq \frac{1}{8}
 \label{82}
\ea
and the bond saturates whenever the following BPS first order eqs. hold
\ba
\vec F_{ij} + \frac{\alpha}{1 + \alpha^2} \varepsilon_{ij}(1 - \Phi^2) \vec \psi &=& 0
\nonumber\\
 D_i\vec \Phi
+ \alpha \varepsilon_{ij} \vec \Psi
 \wedge
{D_j\vec\Psi} &=& 0
\ea
Now, one can easily see that these two eqs. are compatible if and only if $\alpha = \pm 1$
this implying
\be
\lambda = \frac{e^2}{8\pi} = 1
\ee
and we again find the Bogomol'nyi point as a necessary condition for bounding the energy.

In summary, the complete set of BPS equations can be written as
\ba
\vec F_{ij} \pm \frac{1}{2} \varepsilon_{ij}(1 - \Phi^2) \vec \psi &=& 0
\nonumber\\
 D_i\vec \Phi
\pm  \varepsilon_{ij} \vec \Psi
 \wedge
{D_j\vec\Psi} &=& 0\nonumber\\
D_i\Psi &=& 0
\ea
with the double sign related to the  $n$ sign. At the bound, the energy is again proportional
to the magnetic flux
\be
E = \Phi_0^2 \pi |n|
\ee
but, as explained above, $|n|$ is not the topological charge. Indeed, the topological
charge is, in this case, an integer modulo 2 and hence the more stable solutions other than
the trivial vacuum are those with $n = \pm 1$.

\section{ Nonabelian Chern-Simons vortices}
Non-Abelian Chern-Simons vortices have been first discussed in
\cite{deVegS1}-\cite{deVegS2}. Let us start by writing the Chern-Simons
action for the case of an $SU(N)$ gauge theory in $d=3$ space-time dimensions,
\be
S_{CS}[A] = \frac{\mu}{2} \varepsilon^{\alpha\beta\gamma} {\rm Tr}
\int d^3x
\left(
F_{\alpha\beta} A_\gamma -\frac{2}{3} A_\alpha A_\beta A_\gamma
\right)
\ee
This is not a gauge invariant action. Indeed, under a gauge transformation
\be
A_\mu^g = g^{-1} A_\mu g + \frac{i}{e} g^{-1} \partial_\mu g
\ee the action changes according to
\be
S_{CS}[A^g] = S_{CS}[A] + 8\pi^2 \mu  \omega[g]
\ee
with $\omega[g]$ the winding number associated to the gauge transformation,
\be
\omega[g] = \frac{1}{24 \pi^2} {\rm Tr} \varepsilon^{\alpha\beta\gamma}
\int g^{-1} \partial_\alpha g g^{-1} \partial_\beta g g^{-1} \partial_\gamma g
= m \in \mathbb{Z}
\ee
Since for quantizing the theory one needs that
\be
\exp(iS_{CS}[A^g]) = \exp(iS_{CS}[A])
\ee
(with $\hbar =1$)
one   needs  the coefficient $\mu$ to be quantized
\be
\mu = \frac{e^2m}{4\pi} \; , \;\;\; m   \in \mathbb{Z}
\label{qondi}
\ee
Note that this condition is valid both in Minkowski and in Euclidean space since
the CS term does not depend on the metric and is then
unaffected by a Wick rotation.

We can then add the Chern-Simons term
to the Yang-Mills Higgs system (\ref{1na}) (considered
 in $2+1$ dimensions)
\ba
S &=& {\rm Tr}  \int d^3x
\left( -\frac{1}{2} F_{\mu\nu}F^{\mu\nu} +
\sum_{A=1}^{N-1} D_{\mu}\Phi^A D^{\mu}\Phi^A
+ D^\mu\Psi D_\mu\Psi  \right.\nonumber\\
&-& \left. V(\Phi,\Psi)
\vphantom{\sum_{A=1}^{N-1}}\right) + S_{CS}[A]
 \label{1na2}
\end{eqnarray}
and look for vortex like solutions which, as in the Abelian case, should now carry not
only magnetic but also electric flux.

We shall again propose an
axially symmetric ansatz as (\ref{super}) but this time including also the $A_0$
field,

\begin{eqnarray}
\Phi^A &=& f^a (r)
\Omega_n^{-1}(\varphi)(E_{a_A} + E_{a_A}) \Omega_n(\varphi) \; , \;\;\; A=1,2,\ldots, N-1
\nonumber\\
\Psi &=& \sum_{j=1}^{N-1}C_j H_j\nonumber\\
A_\varphi  &=& \frac{n}{e}\, a(r) \mathbf{M} \; , \;\;\; A_r = 0
\nonumber\\
A_0 &=& \frac{n}{e} a_0(r) \mathbf{M}
\nonumber\\
\mathbf{M} &=&  {\rm diag} (1/N,1/N, \ldots, 1/N, (1-N)/N)
\label{super2}
\end{eqnarray}
Given the ``electromagnetic tensor'' defined as in (\ref{ele})
\be
{\cal F}_{\mu\nu} = \frac{{\rm Tr}
(\mathbf{M}  F_{\mu\nu})}{{\rm Tr} (\mathbf{M}^2)}
\label{ele2}
\ee
with a magnetic flux defined as in (\ref{flujo}),
\be
\Theta = \frac{1}{2} \int d^2x \varepsilon_{ij}{\cal F}_{ij}=
 \frac{2\pi }{\sqrt Ne} n
\label{flujo2}
\ee
we have now an electric charge $Q$ which can be defined as follows.
We first define the electric field from (\ref{ele2}) as
\be
 E_i = {\cal F}_{oi}
 \ee
 which can be seen to reduce, with our axially symmetric ansatz to
 \be
 E_r = {\cal F}_{0r} = -\frac{1}{\sqrt N} \frac{da_0 }{dr}
 \ee
 Then, from the Gauss law,
 \be
 \frac{dE_r}{dr} + \sigma  = \mu B
 \label{gaus22}
 \ee
 where
 \be
B =  \frac{1}{2}   \varepsilon_{ij}{\cal F}_{ij}  \; , \;\;\; \sigma  =
e^2 f^2  a_0
\ee
since $\lim_{r \to \infty} E_r = 0 $ according to the boundary conditions,
one gets, from (\ref{gaus22}) a relation between the charge $Q$,
\be
Q = \int d^2x  \sigma
\ee
and the flux,
\be
Q = \mu \Phi
\label{fluxi}
\ee
Now, due to the quantization condition (\ref{qondi}), eq.(\ref{fluxi})
becomes
\ba
Q &=& mn Q_0 \; ,    \;\;\;  m,n  \;\;\; \in \mathbb{Z}\nonumber\\
Q_0 &=& \frac{e}{2\sqrt N}
\ea
Charge quantization can be connected with the angular momentum
J of the vortex,
\be
J = \int d^2x \varepsilon_{ij} x_i T_{oj}
\ee
with $T_{\mu\nu}$ the energy momentum tensor. One can easily find that
\be
J = -\frac{2Q}{\sqrt N e} = - \frac{1}{2N} nm
\ee

Starting from ansatz (\ref{super2}) solutions can be constructed
exactly as in the previous cases. Both the magnetic field and
the Higgs scalar $\Phi$ have the same qualitative behavior as
for neutral vortices. Concerning the electric field, it can be related
with that of the abelian Chern-Simons  model already discussed.

Also as in the Abelian case, one can find Bogomol'nyi equations and a bound starting
from a Lagrangian where  the Yang-Mills term is absent and choosing a sixth-order
potential for the $\Phi$ field \cite{CLMS}. restricting to the $SU(2)$ case
one starts from

\be
S =  S_{CS}[A] + {\rm Tr}  \int d^4x
\left(
  D_{\mu}\Phi^a D^{\mu}\Phi^a
+ D^\mu\Psi D_\mu\Psi - V(\Phi,\Psi)
\right)
 \label{1na3}
\ee
with
\be
V(\Phi,\Psi) = V_1[\Phi^a\Phi^a] + V_2[\Psi^a\Psi^a] + g (\Psi^a\Phi^a)^2
\ee
Again the $\Psi$ field will be taken as an spectator just present to completely break
the symmetry. It will be chosen so as to make vanish $V_2$. The third term
precisely ensures symmetry breaking. Concerning $V_1$, as in the Abelian case, if one
wants to obtain first order Bogomol'nyi equations one needs a sixth order potential
with a coupling constant which is related to the gauge coupling constant
\be
V_1[\Phi] = \frac{e^4}{8\mu^2} \Phi^a\Phi^a \left(\Phi^a\Phi^a - \Phi_0^2\right)^2
\ee

The energy for a static configuration is given by
\be
E = \int d^2x T_{00} = \int d^2x \left(
\frac{e^2}{2} (\vec A_0 \wedge \vec \Phi)^2 + \frac{1}{2}( D_i\vec\Phi)^2
+ V_1[|\vec \Phi|)
\right)
\ee
Now the Gauss law,
\be
e(\vec \Phi \wedge D^0 \vec \Phi) = \frac{\mu}{2} \varepsilon_{ij} \vec F_{ij}
\equiv \mu \vec B
\ee
allows to write
\be
\vec A_0 \wedge \vec \Phi = \frac{\mu}{e^2|\Phi|^2}\vec B \wedge \vec \Phi
\ee
so that the energy becomes
\be
E = \int d^2x \left(
   \frac{\mu^2}{2e^2|\Phi|^2}|\vec B|^2   + \frac{1}{2}( D_i\vec\Phi)^2
+ V_1[|\vec \Phi|)
\right)
\ee
which can be written a la Bogomol'nyi in the form
\ba
E &=&  \int d^2x \left(
\frac{1}{4}
(D_i\vec \Phi \pm \varepsilon_{ij}\check{\Psi}\wedge D_j\vec\Phi)^2 +
\right. \nonumber\\
&&  \left. \frac{\mu^2}{2e^2|\Phi|^2}(B \mp \frac{e^3}{2\mu^2} |\Phi|^2
(|\Phi|^2 - \Phi_0^2)\check \Psi)^2 \mp \frac{e}{2} \Phi_0^2
(\check \Psi . \vec \Phi)
\right)
\ea
We recognize in the last term the magnetic flux
\be
\Theta = \frac{1}{2} \int d^2x \check \Psi . \vec B =
 \frac{2\pi }{e} n
\label{flujo3}
\ee
so that the energy is again bounded by the flux,
\be
E \geq \Phi_0^2 \pi n
\ee
and the bound is attained whenever the first order BPS equations hold,
\ba
D_i\vec \Phi \pm \varepsilon_{ij}\check{\Psi}\wedge D_j\vec\Phi &=& 0 \nonumber\\
 \vec B \mp \frac{e^3}{2\mu^2} |\Phi|^2
(|\Phi|^2 - \Phi_0^2)\check \Psi &=&0 \nonumber\\
D_i\vec \Psi &=& 0
\ea
Qualitatively, the vortex solutions to this system are the same as those
solving the second-order Euler-Lagrange equations described in the precedent
section.

\section*{More on non-Abelian vortices and \\semi\-lo\-cal strings}

We have already constructed semilocal vortices for the case in which the gauge group
was $U(1)_{local}$ but one considers an $SU(2)_{global}$ doublet of complex scalars.
One can  extend the construction  to a non-Abelian gauge group $U(N)_{local}$ provided
the global group $SU(N_f)_{global}$ is such that $N_f> N$.

Before explaining this,  we shall start buy
describing a class of non-Abelian vortices different from those
discussed in sections 5 and 6. They were originally discussed in
refs.\cite{AB}-\cite{Mar}.
Let us consider gauge fields taking values in an
$U(N_c)_{gauge}$ theory and scalars $\phi^a_i$ carrying not only a color index
 $a=1,2, \ldots,N$
but also a flavor index $i=1,2, \ldots, N_f$. We shall first consider $N=N_f$ and
write these scalars as an $N \times N$ matrix,
\be
\Phi = (\phi^a_i)  \; , \;\;\; a,i=1,2,\ldots, N
\ee
The action is
\ba
S&=&  \int d^4x \left(  \frac{1}{4e^2} {\rm Tr}(F^{\mu\nu}F_{\mu\nu})
+ {\rm Tr}(D_\mu\Phi^\dagger D^\mu\Phi)
- \frac{e^2}{8} {\rm Tr}(\Phi^\dagger T^A \Phi)^2
 \right. \nonumber\\
 && \left. - \frac{e^2}{8}({\rm Tr}(\Phi^\dagger\Phi) - \Phi_0^2)^2
\right)
\label{actionx}
\ea
Here $T^A \; , A=1,2,\ldots, N$ are the $SU(N)$ generators normalized
so that ${\rm Tr} (T^AT^B) = 1$ and
\be
D_\mu\Phi = (\partial_\mu -\frac{i}{2} A_\mu^0 - \frac{i}{2} A^A_\mu T^A)\Phi
\label{cov}
\ee
The theory has a $U(N)_{local}\times SU(N)_{global}$  symmetry. On the scalars this acts
according to
\be
\Phi \to U \Phi V \; , \;\;\; U \in U(N)_{local} \;\;\; V\in SU(N)_{global}
\ee

Concerning symmetry breaking, the last term in (\ref{actionx}) forces $\Phi$ to
develop a vacuum expectation value
while the last but one forces it to diagonal,
\be
\Phi_{vac} = \Phi_0 \delta^a_i
\label{vacc}
\ee
This vev is preserved only for transformations in which $U = V^{-1}$ that is, one performs
a global gauge transformation and a related global flavor transformation simultaneously.
The pattern of spontaneous symmetry breaking is then
\be
U(N)_{local}\times SU(N)_{global} \to SU(N)_{diag}
\ee
This mechanism was known as  {\it color-flavor locking}
in a QCD context, \cite{BH}-\cite{W}.

Let us now discuss the vortex solutions of this model. For simplicity,
 we shall
 consider the case $N=2$ \cite{SY} but the generalization to $SU(N)$ will be
 evident.

 Since there is a $U(1)$
gauge symmetry which is spontaneously broken, there are of course Abelian Nielsen-Olesen
vortex solutions which we have already studied. But there are other novel solutions,
related to the non-Abelian character of the theory
which are obtained as follows.

Let us take the vacuum (\ref{vacc}) and make just one of its components,
 to wind. This, in the particular $N=2$ case means
\begin{equation}\label{117}
\Phi_v  = \Phi_0\left (\matrix{ \exp(i\varphi) &   0\cr
0 & 1 }\right )
\end{equation}

Once one has such a vacuum, it is easy to write an ansatz for the Higgs scalar everywhere,
\begin{equation}\label{1176}
\Phi  = \Phi_0\left (\matrix{ f_1(r)\exp(i\varphi) &   0\cr
0 & f_2(r) }\right )
\end{equation}
such that

\[
 \lim_{r\to \infty} f_1(r) = \lim_{r\to \infty} f_2(r) = \Phi_0
 \]
 \be
 f_1(0) = f_2(0) = 0
\ee
Concerning the ansatz for the gauge field,

\ba
A_i^{SU(N)} &=& (1 - a_3(r)) \partial_i \varphi \left(\matrix{ 1 &   ~~0\cr
0 & -1 }\right )
\nonumber\\
A_i^{U(1)} &=&  (1-a(r))
\partial_i \varphi
\ea
with
\[
 \lim_{r\to \infty} a(r) = \lim_{r\to \infty} a_N(r) =  0
 \]
 \be
 a(0) = a_N(0) = 1
\ee
This ansatz, in which ${\Phi_v}_{11}$ is the scalar components that winds is usually called
a $(1,0)$ string. If one chooses ${\Phi_v}_{22}$ to be
the sole  topologically nontrivial component
will have a $(0,1)$ string,
\begin{equation}\label{1177}
\Phi_v  = \Phi_0\left (\matrix{ f_1(r) &   0\cr
0 & f_2(r)\exp(i\varphi) }\right )
\end{equation}

Let us note that if we define
\be
\Omega_i[\Phi] = \frac{1}{i}\left(\partial_i \Phi\right) \Phi^{-1}
\ee
then, for the $(1,0)$ string one has
\be
\Omega_i[\Phi_v] =  \left(\matrix{ 1 &   0\cr
0 & 0 }\right )  \partial_i\varphi
\ee
Then, if we associate $\Omega_i$ with a topological charge $T$ through
the formula
\be
T = {\rm Tr} \oint \Omega_i dx^i
\ee
so that
\be
T_{(1,0)} = T_{(0,1)} = 2\pi
\ee
Let us note that if instead of vacuum (\ref{117}) or (\ref{1177})
one proposes  an ``Abelianized'' one
\begin{equation}\label{1178}
\Phi_v  = \Phi_0\left (\matrix{ \exp(i\varphi) &   0\cr
0 &   \exp(i\varphi) }\right) = \exp(i\varphi) I
\end{equation}
 one would obtain for T
 \be
 T_{(1,1)} = 4\pi = T_{(1,0)} + T_{(0,1)}
 \ee
Now, since the covariant derivative (\ref{cov}) has to vanish at infinity,
\be
\left. D_i\Phi \right|_\infty =
\left(\partial_i -\frac{i}{2} A_i^0 - \frac{i}{2} A^A_i T^A\right)
\left. \Phi_v \right|_\infty= 0
\ee
one easily finds that
\be
 \frac{1}{i}  (\partial_i \Phi_v) \Phi_v^{-1}  = \frac{1}{2}
 \left.\left( A_i^0 +   A^A_i T^A\right)\right|_\infty
\ee
\be
T =\frac{1}{2} {\rm Tr} \oint (A_i + A_i^A T^A)
dx^i =  \oint A_i d x^i
\ee
Note that only the $U(1)$ component of the gauge field
contributes to the topological charge
We  then conclude that   the magnetic flux $\Phi$ of these vortex solutions
is precisely connected to T,
\be
\Omega = \frac{1}{e} T
\ee
and in accordance, one can see that the Bogomol'nyi bound will turn to be
\be
E \geq \Phi_0 \Omega =  \frac{1}{e}\Phi_0 T
\ee

\section{Bogomol'nyi equations and Super- \\
sym\-metry algebra}
We already mentioned at the beginning that the Bogomol'nyi point
(\ref{bog}),
 \be
 e^2 = 8 \lambda
 \label{bog222}
 \ee
at which solutions to a first-order set of equations (Bogomol'nyi
eqs.) exist and an exact vortex solution is known, not only
corresponds to the limiting point separating type-I from type-II
superconductivity, but, as noticed in \cite{dVS}, coincides with
that which has to be chosen in order to have a supersymmetric
extension of the model

This connection between Bogomol'nyi equations and bounds and
supersymmetry began to be clarified after the work of Witten and
Olive \cite{WO}.
Let us then review the results in that paper which
starts with a two $1+1$ dimensional supersymmetric model with Lagrangian
\be L=
- \frac{1}{2} \left(  (\partial_\mu \phi)^2 +
 \bar \psi \not \!\partial \psi +  V^2[\phi] +
  V'[\phi] \bar \psi \psi \right) \ee where $\psi$ is a
Majorana fermion and $V[\phi]$ an arbitrary function.
Our metric is ${\rm diag}\, g_{\mu\nu} = (-1,1)$,  ($\mu=0,1$).  The
 gamma matrices are
\be
\gamma^0 = - i \sigma^2  = \left (\matrix{ 0 &   -1\cr
1 &  ~0 }\right )   \; , \;\;\;
\gamma^1 =  \sigma^3  = \left (\matrix{ 1 &  ~0\cr
0 &  -1 }\right )
\ee
The charge conjugation matrix $C$ is
\be
C = \sigma^2 \; , \;\;\; (\gamma^\mu)^{\rm t} C = - C \gamma^\mu
\ee
Here $\sigma^i$ are the standard Pauli matrices.
We write the Majorana spinor in the form
\be
\psi =  \left(\matrix{ \psi^+  \cr
\psi^-  }\right )
\ee
and then
\be
\bar \psi =  \psi^{\rm t} C = \left(\psi^-  -\!\psi^+
\right)
\ee

The equations of motion are,
\ba
&& \Box \phi = V V' + \frac 12 V'' \bar \psi \psi \nonumber\\
&& \db \psi - V' \psi = 0
\label{eqsmot}
\ea

Under the following supersymmetry transformations
\ba
\delta \phi &=&  \bar \epsilon \psi\nonumber\\
\delta \psi &=& \left( \db \phi
- V[\phi] \right) \epsilon \nonumber\\
\delta \bar \psi &=& -\bar \epsilon \left(\db \phi
 + V[\phi] \right)
\label{varia}
\ea
the Lagrangian changes as a total divergence and then one finds,
using Noether theorem that the conserved supersymmetry current is
\be
J^\mu = -\left( \db \phi+ V[\phi]\right)\gamma^\mu \psi
\ee
so that the conserved supercharge is
\be
Q = -\int dx \left( \db \phi + V[\phi]\right)\gamma^0 \psi
\label{sicu}
\ee
Writing
\be Q =  \left(\matrix{ Q^+  \cr
Q^-  }\right )
\ee
one  finds from (\ref{sicu})
\ba
Q^+ &=&\int dx
\left(\partial_0 \phi\, \psi^+ + (\partial_1 \phi+ V[\phi]) \psi^-
\right)
\nonumber\\
Q^- &=&\int dx \left(
\partial_0 \phi \,  \psi^- -( \partial_1 \phi - V[\phi]) \psi^+\right)
\label{cargas}
\ea

Calling $H$ the Hamiltonian and $Z$ the quantity
\be
Z =   \int dx V[\phi] \frac{\partial \phi}{\partial x}
\ee
one finds
\be
\{ Q^\pm,Q^\pm\} = 2 (H \pm Z)
\label{qq}
\ee
where we have used
\be
\{\psi^\pm(x,t),\psi^\pm(cy,t)\} = \delta (x-y)
\ee
Note that $Z$ can be written in the form
\be
 Z =  \int dx V[\phi] \frac{\partial \phi}{\partial x} =  \int dx
 \frac{\partial  F[\phi]}{\partial x}
 \ee
 with $(\partial F[\phi])/(\partial \phi) = V[\phi]$. Then, taking as an
 example the usual symmetry breaking potential with two ground states $\phi= \pm a$,
 \be
 V = -(\phi^2 - a^2)
 \ee
 one has
 \be
 F = \lambda(a^2\phi - \frac{1}{3} \phi^3)
 \ee
and so
\be
Z = \lambda \int_{-\infty}^{\infty} dx
\frac{\partial}{\partial x}\left( a^2  \phi - \frac{1}{3} \phi^3
\right)
\ee
Consider static configurations. This integral vanishes for topologically trivial states,
$\phi(x= -\infty) = \phi(x=\infty) = \pm a$. If instead one takes
 $\phi(x= -\infty) = \mp a$ and $\phi(x= \infty) = \pm a$  one has
$ Z = \pm 4\lambda a^3/3 $. One associates the positive value
for a kink state and the negative value for anti-kink state.
Appropriately normalized, the value of
$T$ can be related to the usual
kink topological charge defined as
\be
 T = \frac{1}{2a} \int_{-\infty}^{\infty}
 dx \frac{\partial \phi}{\partial x} = \phi(x= \infty) - \phi(x= -\infty)
\ee

Eq.(\ref{qq}) implies, for a particle with rest mass $M=H$ that
\be
M \geq  |Z|
\label{boundbound}
\ee
and the bound is saturated for those states $|phys\rangle$ such that
\be
Q_+ |phys\rangle = 0    \;\;\; {\rm or } \;\;\;  Q_-|phys\rangle = 0
\label{158}
\ee
so that physical soliton or anti-soliton states such that
 (\ref{or}) holds
 are annihilated by $Q_+$ or $Q_-$. Now,
eq.(\ref{158}) implies, according to (\ref{cargas}) (in
the static case),
\be
(\partial_1\phi + V[\phi])  \psi_- = 0 \;\;\; {\rm or } \;\;\;
(\partial_1\phi - V[\phi]) \psi_+  = 0
\label{or}
\ee
and then one has the following Bogomol'nyi eqs.
\be
\partial_1\phi = - V[\phi]  \;\;\; {\rm or } \;\;\;
\partial_1\phi =  V[\phi]
\label{bogo1}
 \ee
Let us note at this point that the energy for static
configurations (the rest mass M) in the
bosonic sector takes the form
\be
M = \int dx \frac{1}{2}  \left(\partial_1\phi)^2  + V^2[\phi]
\right)
\ee
and can be trivially written a la Bogomol´nyi in the form
\be
M = \int dx \frac{1}{2}  \left( (\partial_1\phi)  \pm  V[\phi]
\right)^2  \mp Z
\label{ecuac}
\ee
from which the  bound (\ref{boundbound}) can be equally obtained.

Of course, the solution to these BPS equations also solve the second order
Euler-Lagrange eqs. Moreover, one can look at  eq.(\ref{bogo1}) as the
square root of the equation of motion for static bosonic  configurations.
One can see this by integrating the equation for configurations such that
$\partial\phi/\partial x$ and $V[\phi]$ vanish at infinity
\be
\partial_1^2 \phi  =  V V'  \; \;\;\;  \rightarrow   \; \;\;\;
(\partial_1 \phi)^2 =
V[\phi]^2
\ee

Note that since the fermionic supersymmetry transformation law (\ref{varia})
can be written, for static configurations, as
\be
\delta \psi = %
  \left (\matrix{ \partial_1\phi - V &   0\cr
0 &  -\partial_1\phi - V }\right )
  \left(\matrix{ \epsilon^+  \cr
\epsilon^-  }\right )
\ee
putting this variation to zero is an alternative way to obtain
the Bogomol'nyi eqs without necessity of constructing
the supercurrent.

One can easily see that there is a zero mode for the Dirac equation
associated with each Bogomol'nyi equation. These zero modes are
\be
\chi_0 = (\db \phi - V)  \left(\matrix{ \chi^+_0  \cr
0  }\right )  \;\;\; {\rm or} \;\;\;
\chi_0 = (\db \phi + V)  \left(\matrix{0 \cr
 \chi^-_0  }\right )
 \label{167}
 \ee
where $\chi^\pm_0$ are constant components. One can understand this
by noting that the kink (antikink) solution
of the Bogomol'nyi eqs.(\ref{bogo1}) solves, together with $\psi = 0$, the
coupled Euler-Lagrange
eqs. of the model.  But a supersymmetric transformation of
such solution will
still be a solution and this means that $\delta \psi$ (the SUSY transform of $\psi =0$)
will solve the Dirac equation. Such $\delta \psi$ is precisely ({167}).

Let us end by stressing that  eqs.(\ref{qq}) correspond
to an $N=2$ extended supersymmetry
algebra with central charge Z; the equality   $M = Z$
when Bogomol'nyi bound is saturated eliminates half
of the SUSY generators
 (either $Q_-$ or $Q_+$) which in turn is associated with a
 Dirac equation zero-mode.

\section*{Supersymmetry and ANO vortices}
We are now ready to see how the vortex BPS equations for the
Abelian Higgs
model can be derived by SUSY considerations. We shall follow here
refs.\cite{LLW}-\cite{ENS}. One has here two possibilities; either one studies
the $N=2$ supersymmetric extension of the $d= 2+1$-dimensional model looking
at vortices as
static solutions of eqs.~of motion or one analyzes the $2$-dimensional
(Euclidean) model in which vortices can be seen as instanton solutions. Let us start
by discussing the $d=2+1$ approach.

\subsection*{The $2+1$ $N=2$ supersymmetric model}
We take as metric ${\rm diag} g_{\mu\nu} = (+--)$. A two component Majorana spinor
will be written as
\be
\left(\psi^\alpha\right) =  \left(\matrix{ \psi^+  \cr
\psi^-  }\right )
\ee
Spinorial indices are raised and lowered  using
  $C_{\alpha\beta} = i \varepsilon_{\alpha\beta}$ with $\varepsilon_{-+} = 1$,
\be
\psi^\alpha = C^{\alpha\beta}\psi_\beta
\ee
Dirac matrices are chosen as
\be
\gamma^1 = \sigma_1 \; , \;\;\; \gamma^2 = \sigma_2 \; , \;\;\;
\gamma^0 = \sigma_3 \; , \;\;\;
\ee
with $\sigma^i$ the standard Pauli matrices. Superspace coordinates
are written as
$(x^\mu;\theta^\alpha)$ with $\theta^\alpha$ two anticommuting
spinor coordinates.

The Higgs field $\phi$, together with a higgsino $\psi$ and an
auxiliary field $F$ are accommodated in a complex scalar superfield,
\be
\Phi(x;\theta) = \phi(x) + \bar\theta\psi(x) -
\frac{1}{2}\bar\theta \theta F(x)
\ee
The gauge field $A_\mu$ together with a photino $\rho$ are accommodated
in a vector superfield
\be
\Gamma_\alpha(x;\theta) = i A_\mu(x) (\gamma^\mu\theta)_\alpha -
\bar\theta \theta \rho_\alpha(x)
\ee
Finally a real scalar superfield will be necessary in order
to implement the Fayet-Iliopoulos gauge symmetry breaking mechanism,
\be
S(x;\theta)= N(x) +
 \bar \theta \chi(x) - \frac{1}{2}\bar\theta \theta D(x)
\ee
where $N$ is a scalar, $\chi$ is a Majorana fermion and
$D$ an auxiliary field.

Calling ${\cal D}_\alpha$ the (super) covariant derivative,
\be
{\cal D}_\alpha = \frac{\partial}{\partial \theta^\alpha}
+ i (\gamma_\mu \theta)_\alpha
\partial_\mu
\ee
the spinorial electromagnetic superfield $W_\alpha$ is given by
\be
W_\alpha = \frac{1}{2}  {\cal D}_\beta{\cal D}_\alpha \Gamma^\beta
= \rho_\alpha(x) + \frac i2 \varepsilon_{\mu\nu\sigma}F^{\nu\sigma}(x)
(\gamma^\mu\theta)_\alpha - \frac i2 \bar\theta \theta
(\db \rho)_\alpha(x)
\ee
With all this the $N=1$ supersymmetric version of the Abelian Higgs
model Lagrangian takes, in terms of superfields, the form
\ba
L_{N=1} &=& \int d^2\theta \left(
\frac 12 W^\alpha W_\alpha
-\frac 14 \left( {\cal D}^\alpha +ie \Gamma^\alpha\right) \Phi^*
\left( {\cal D}_\alpha -ie \Gamma_\alpha\right) \Phi \right.
\nonumber\\
&& \left. -\frac 14 D^\alpha S D_\alpha S + \sqrt{2 \lambda} S \Phi^*\Phi
+ \eta S
\right)
\ea
where $\eta$ is the real Fayet-Iliopoulos coefficient. Integration out over
$\theta$ one gets the $N=1$ supersymmetric Lagrangian in components which
takes the form
\ba
L_{N=1} &=& -\frac14 F^{\mu\nu}F_{\mu\nu} +
 \frac 12 \partial^\mu N\partial_\mu N +
 \frac12 (D^\mu\phi)^* D_\mu\phi +\frac 12 D^2
 \nonumber\\
 &&+ \sqrt{2\lambda} D  \vert \phi\vert^2 +
 \eta D + \frac 12  \vert F\vert^2 +
 \sqrt{2\lambda} N(F^*\phi + F\phi^*)\nonumber\\
 && \frac i2 \bar \rho\!\db \rho + \frac i2 \bar \chi \!\db \chi
 + \frac i2 \bar \psi \not\!\!D \psi - \sqrt{2\lambda} N \bar \psi \psi
 \nonumber\\
&& +\frac{ie}2 (\bar\psi \rho\phi - \bar \rho \psi\phi^*)
 -\sqrt{2\lambda}(\bar \psi \chi \phi + \bar \chi \psi \phi^*)
 \label{lagg}
\ea
Solving the equations of motion for the auxiliary fields one gets
\be
D = \sqrt{2\lambda}|\phi|^2 + \xi    \; , \;\;\; F = -2\sqrt{2\lambda}N\phi
\ee
Here $\xi$ is the Fayet-Ilopoulos parameter which will be written as
\be
\xi = -\sqrt{2\lambda}\phi_0^2
\ee
so that symmetry breaking takes place for
\be
|\phi| =  \phi_0 > 0
\ee
With this, Lagrangian (\ref{lagg}) takes the form
\ba
L_{N=1} &=& -\frac14 F^{\mu\nu}F_{\mu\nu} +
 \frac 12 \partial^\mu N\partial_\mu N +
 \frac12 (D^\mu\phi)^* D_\mu\phi -4\lambda N^2 |\phi|^2
 \nonumber\\
&& - \lambda(|\phi|^2 - \phi_0^2)
+ \frac i2 \bar \rho\!\db \rho + \frac i2 \bar \chi \!\db \chi
 + \frac i2 \bar \psi \not\!\!D \psi - \sqrt{2\lambda} N \bar \psi \psi
 \nonumber\\
 && + \frac{ie}2 (\bar\psi \rho\phi - \bar \rho \psi\phi^*)
 -\sqrt{2\lambda}(\bar \psi \chi \phi + \bar \chi \psi \phi^*)
 \label{lagg2}
\ea
This Lagrangian is invariant under the following infinitesimal
supersymmetry transformations
\be
\delta\rho =
 -\frac{i}{2}\varepsilon_{\mu\nu\alpha}F^{\nu\alpha}\gamma^\mu \epsilon \; , \;\;\;
\delta A_\mu = -i\bar \epsilon \gamma_\mu \rho
\label{eq1}
\ee
\be
\delta N = \bar \epsilon\chi\; , \;\;\;
\delta \chi = -\sqrt{2\lambda}(|\phi|^2 - \phi_0^2)\epsilon -
i\gamma^\mu\epsilon \partial_\mu N
\label{eq2}
\ee
\be
\delta\phi = \bar \epsilon \psi\; , \;\;\; \delta\psi = -i\gamma^\mu\epsilon
D_\mu\phi
-\sqrt{8\lambda}N\phi\epsilon
\label{eq3}
\ee
with $\epsilon$ an anticommuting real (i.e. Majorana spinor) parameter.

It is possible to extend this $N=1$ supersymmetry to an $N=2$ supersymmetry. This
can be achieved by considering SUSY transformations (\ref{eq1})-(\ref{eq3}) where
the parameter $\epsilon$ is replaced by a complex one, $\epsilon_c$ and real
fermions $\chi$ and $\rho$   are combined into a Dirac fermion $\Sigma$,
\be
\Sigma = \chi - i \rho
\label{transfS}
\ee
In terms of this new field the Lagrangian (\ref{lagg2}) becomes
\ba
L_{N=2} &=& -\frac14 F^{\mu\nu}F_{\mu\nu} +
 \frac 12 \partial^\mu N\partial_\mu N +
 \frac12 (D^\mu\phi)^* D_\mu\phi -4\lambda N^2 |\phi|^2
 \nonumber\\
&& - \lambda(|\phi|^2 - \phi_0^2)
+ \frac i2 \bar\Sigma \!\db \Sigma
 + \frac i2 \bar \psi \not\!\!D \psi - \sqrt{2\lambda} N \bar \psi \psi
 \nonumber\\
 && - \frac{e + \sqrt{8\lambda}}{4} (\bar \Sigma \phi + h.c.)
 + \frac{e - \sqrt{8\lambda}}{4} (\bar \Sigma^c \phi + h.c.)
 \label{lagg3}
\ea
where $\Sigma^c$ is the charge conjugate (the complex conjugate) of $\Sigma$.

Concerning the SUSY transformations (\ref{eq1})-(\ref{eq3}), they take the
form, in terms of $\Sigma$,
\be
\hat\delta\Sigma =
-\left(\frac{1}{2}\varepsilon_{\mu\nu\alpha}F^{\nu\alpha}\gamma^\mu
+ \sqrt{2\lambda}(|\phi|^2 - \phi_0^2) + i\!\db N\right)\epsilon^c
\ee
\be
\hat \delta A_\mu =  \frac{1}{2} \left(
\bar\epsilon^c \gamma_\mu \Sigma + h.c.
\right)
\; , \;\;\;
\hat \delta N = \frac{1}{2}\left(\bar\epsilon^c\Sigma + h.c.
\right); , \;\;\;
\hat \delta\phi = \bar \epsilon^c \psi
\ee
\be
 \hat \delta\psi =-\left(i\gamma^\mu
D_\mu\phi
+\sqrt{8\lambda}N\phi\right)\epsilon^c
\label{eq33}
\ee
Here $\epsilon^c$ is a complex parameter which can be written as
\be
\epsilon^c = \epsilon \exp(i\alpha)
\ee
and so  (\ref{eq33}) can be interpreted as
transformations (\ref{eq1})-(\ref{eq3}) with real $\epsilon$
followed by a phase
transformation for fermions,
\be
\Sigma \to \exp(i\alpha)\Sigma \; , \;\;\; \psi \to \exp(i\alpha)\psi
\ee
However the last   term in (\ref{lagg3}) is not invariant under this phase rotation,
\be
\bar \psi \Sigma^c \phi + h.c. \to  \exp(2i\alpha) \bar \psi \Sigma^c \phi + h.c.
\ee
so one needs
\be \lambda = \frac{e^2}{8} \label{finalmente} \ee as necessary and
sufficient condition for the $N=2$ SUSY. This was a well-known
result holding in general when, starting from an $N=1$
supersymmetric gauge model, one attempts to impose a second SUSY
(see for example \cite{Sohnius}): conditions on coupling constants
have to be imposed so as to accommodate different $N=1$ multiplets
into an $N=2$ multiplet. Eq.(\ref{finalmente}) is an example of such
conditions. These kind of conditions for the Abelian Higgs model
were obtained, following
 different routes,  in
\cite{Fayet}-\cite{SalSt},\cite{dVF}.

We shall now study the supersymmetry algebra in order to reobtain
the results above in a way closer to the one used by Olive and
Witten
for their 1+1 model.

The Noether current associated with supersymmetry invariance can be  compactly
written in the form
\be
 {\cal J}^\mu = \sum_{\Phi,\Psi}
\left(\frac{\delta L}{\delta\nabla_\mu\Phi}\delta\Phi
+
 \frac{\delta L}{\delta\nabla_\mu\Psi}\delta\Psi\right) - \Lambda^\mu
\ee
where ${\Phi}$ and ${\Psi}$ represent generic bosonic and
fermionic fields  and  is included to take into account
the possible variation of the
Lagrangian through a divergence term,
\be
\delta L = \partial_\mu \Lambda^\mu
\ee
From this, the conserved supersymmetric charge $Q$ can be defined as
\be
{\cal Q} =   \int d^2 x {\cal J}^0
\ee
 Writing
\be
{\cal Q} = {\bar \epsilon}^c Q + \bar Q \epsilon^c
\ee
in order to take rid of the infinitesimal Grassman parameter
 we find from Lagrangian (\ref{lagg3})
\ba
Q &=& \int d^2x \left(\left(  -\frac12 \varepsilon^{\mu\nu\lambda}
F_{\mu\nu} \gamma_\lambda + i \db N -\frac e2 (\vert \phi\vert^2 - v^2)
\right) \gamma^0\Sigma  + \right.\nonumber\\
&&
\left. \left( i(\not \!\!D \phi)^* - \frac e2 N\phi^*\right)\gamma^0\psi
\right)
\ea
Concerning the conjugate charge $\bar Q$ one has
\ba
\bar Q &=& \int d^2x \bar \Sigma \gamma^0\left(
\left(  -\frac12 \varepsilon^{\mu\nu\lambda}
F_{\mu\nu} \gamma_\lambda - i \db N -\frac e2 (\vert \phi\vert^2 - v^2)
\right)   + \right.\nonumber\\
&&
\left. \bar\psi\gamma^0 \left( -i\not \!\!D \phi - \frac e2 N\phi\right)
\right)
\ea
If we restrict the model to the case $N=0$ and, after computing the Poisson brackets
we put all fermions to zero we end with the original Abelian Higgs model.
Since we are interested in static configurations with $A_0 = 0$ we impose these conditions finding
the following anticommutation relation among spinor supercharges,

\be
\{ Q_\alpha,\bar Q^\beta\}  =  2 \left(\gamma_{0}\right)_{\!\alpha}^{\, \beta}
P^0 + \delta_\alpha^\beta T
\label{anti}
\ee
where
\be
P^0 = E =\int d^2 x \left(\frac14 F_{ij}^2 +\frac12 |D_i\phi|^2 +
  \frac{e^2}{8} (|\phi|^2 - \phi_0^2)^2 \right)
\ee
wile the central charge $T$ is given by
\be
T = \int d^2 x \left(\frac e2 \varepsilon^{ij} F_{ij}
(|\phi|^2 - \phi_0^2) + i \varepsilon^{ij} D_i\phi(D_j\phi)^*\right)
\ee
with $i,j=1,2$.
One can easily see that $T$ can be rewritten in the form
\be
T = \int d^2x \partial_i \Omega^i
\ee
with
\be
\Omega^i = \varepsilon^{ij} \left(e \phi_0^2 A_j + i \phi^* D_j\phi
\right)
\ee
so that using Stokes theorem (and taking into account that $D_i\phi \to 0$ at infinity)
one ends with
\be
T = e \phi^2_0\oint A_i dx^i = e \phi^2_0 \left(\frac{2\pi N}{e}\right)  \; ,
\;\;\; N \in Z
\ee
where the integer $n$ characterizes the homotopy class to which $A_i$ belongs.
To find the Bogomol´nyi bound from the supersymmetry algebra, let us observe that,
being the anticommutators (\ref{anti}) Hermitian one has
\be
\{Q_\alpha,\bar Q^\beta\}\{Q^\alpha,\bar Q_\beta\} \geq 0
\label{boundido}
\ee
Then, using (\ref{anti}) one has
\be
4E^2 - T^2 \geq 0
\ee
or
\be
E \geq \frac{|T|}{2}
\ee
which is nothing but the bound (\ref{boundI}) originally obtained by
Bogomol´nyi by completing squares. The bound is attained for states $|phys\rangle$
which are annihilated by the charges. To extract from this condition the Bogomol´nyi eqs., it
will be convenient to write
\be
Q = \left(\matrix{Q_+ \cr
 Q_-  }\right )   \; , \;\;\;  \bar Q = (\bar Q^+ \bar Q^-)
\ee
after a little work one finds
\ba
Q_+ &=&-\frac12 \varepsilon_{ij}F^{ij} \Sigma_+
-\left(\frac{e}{2}(|\phi|^2 - v^2)\right)\Sigma_-
-i(D_1\phi)^* \psi_+  + (D_2\phi)^* \psi_-
\nonumber\\
Q_- &=&-\frac12 \varepsilon_{ij}F^{ij} \Sigma_-
-\left(\frac{e}{2}(|\phi|^2 - v^2)\right)\Sigma_+
-i(D_1\phi)^* \psi_-  + (D_2\phi)^* \psi_+
\nonumber\\
\ea
Hence, if one combines spinor  components so that
\be
Q_I =( Q_+ +  i Q_-)  \; , \;\;\;
Q _{II} =( Q_+ - i Q_-)
\ee
one has
\ba
Q_I = -\left( \frac12 \varepsilon_{ij}F^{ij}
+ \frac{e}{2}(|\phi|^2 - v^2)\right) \Sigma_I
-\left(i\left(D_1\phi\right)^* - \left(D_2\phi\right)^*
\right)\psi_I \nonumber\\
Q_{II} = -\left( \frac12 \varepsilon_{ij}F^{ij}
- \frac{e}{2}(|\phi|^2 - v^2)\right) \Sigma_{II}
-\left(i\left(D_1\phi\right)^* + \left(D_2\phi\right)^*
\right)\psi_{II}\nonumber\\
\ea
We see that if the bosonic fields satisfy
\ba
&&
\frac12 \varepsilon_{ij}F^{ij}
+ \frac{e}{2}(|\phi|^2 - v^2) = 0\nonumber\\
&&i\left(D_1\phi\right)^* - \left( D_2\phi\right)^* = 0
\label{sib1}
\ea
$Q_I$ vanishes (while $Q_{II}$ not). But if we put
$\epsilon^c_{II} = 0$ we then have ${\cal Q} |phys>=0$
for any configuration satisfying (\ref{sib1}). This in turn will suffice
to ensure that the bound (\ref{boundido}) is attained.

Analogously, the
supersymmetric charge will annihilate physical states if
\ba
&&
\frac12 \varepsilon_{ij}F^{ij}
- \frac{e}{2}(|\phi|^2 - v^2) = 0\nonumber\\
&&i\left(D_1\phi\right)^* + \left( D_2\phi\right)^* = 0
\label{sib2}
\ea
together with $\epsilon^c_{I} = 0$.
But (\ref{sib1}) and (\ref{sib2}) are nothing but Bogomol'nyi
equations so that we see that when the Bogomol'nyi bound is attained
half of the supersymmetry is lost. Again, a fermionic zero mode arises
associated with this supersymmetry breaking and its form can be
just inferred by SUSY transforming a trivial (zero) solution as in the
Witten-Olive example we discussed.

\subsection*{Uniqueness of Bogomol'nyi equations}

I will discuss here how the BPS structure of general gauge theories (depending on the
Maxwell invariant $F^{\mu\nu}F_{\mu\nu}$ and ${\tilde F}^{\mu\nu}F_{\mu\nu}$) coupled
to Higgs scalars is insensitive to the particular form of the gauge Lagrangian.
Indeed,  analyzing their supersymmetric extension, one can
explicitly understand why this happens. Hence Maxwell, Born-Infeld
or more complicated non-polynomial Lagrangians all satisfy the
same Bogomol'nyi equations and bounds which are dictated by the
underlying supersymmetry algebra (\cite{CNS}).

Due to the interest aroused by Born-Infeld theories in the context of the
dynamics of D-branes (see \cite{Tseytlin} and references there)
their Bogomol'nyi equations for different cases were investigated
\cite{G}-
\cite{GNSS} (early
constructions were reported in \cite{NS1}-\cite{NS2}). Bogomol'nyi equations
were found to coincide with those arising in ordinary gauge theories.
Concerning
SUSY
Born-Infeld theories, they
 were originally studied in \cite{DP}-\cite{CF} (see also \cite{H}).

One can quickly understand why BPS equations are not sensitive to the
dynamics that one chooses for the gauge field by noting that they can be derived, in
a SUSY framework, by imposing the
vanishing of (half of) the supersymmetry variations of the gaugino
and higgsino fields and these variations are formally the same for
very different Lagrangians. The dynamics associated with the
Lagrangian enters however through the equation of motion for the
auxiliary field $D$ (of the gauge field supermultiplet) which
appears in the supersymmetric transformation law for the gaugino.
It is then through $D$ that the form of the Lagrangian may in
principle determine the form of the BPS relations.

To see this in detail, let us take as an example  the  Abelian-Higgs  model
 in
$d=3$ dimensions, for which  Bogomol'nyi equations were first derived
\cite{Bog},\cite{dVS} and which we have already discussed in section 2.
The arguments should hold, however, for other models like for
example the $SO(3)$ gauge theory and in other dimensionalities of
space-time.

We shall first consider $d=4$ dimensional Minkowski space  (with
signature $(+,-,-,-)$)  so that one could also use the results to
analyze other $d=4$ models and then proceed to dimensional reduction
to $d=3$. The gauge vector superfield $V$ is written, in the
Wess-Zumino gauge,
\be V = - \theta \sigma^\mu \bar \theta A_\mu + i \theta \theta
\bar \theta \bar \lambda - i \bar \theta \bar \theta \theta
\lambda + \frac{1}{2} \theta  \theta  \bar \theta  \bar \theta D
\label{1x} \ee
Here $A_\mu$ is a vector field, $\lambda = (\lambda_\alpha)$ and
$\bar \lambda = (\bar \lambda^{\dot \alpha})$ are two-component
spinors ($\alpha, \dot \alpha = 1,2$) which can be combined to
give a four-component Majorana fermion and $D$ is an auxiliary
field.

From $V$ the chiral superfield $W_\alpha$ can be constructed,
\begin{equation}
W_\alpha \left( y,\theta ,\bar \theta \right) =-i\lambda _\alpha
+\theta _\alpha D-\frac i2\left( \sigma ^\mu\bar \sigma ^\nu\theta
\right) _\alpha F_{\mu\nu}+\theta \theta \left( \sigma
^\mu\partial _\mu\bar \lambda \right) _\alpha \label{2x}
\end{equation}
Here $\lambda ,$ $\bar \lambda ,$ $D$ and $F_{\mu\nu} =
\partial_\mu A_\nu - \partial_\nu A_\mu$   are functions of the
variable $y^\mu=x^\mu+i\theta \sigma ^\mu\bar \theta $ where
$x^\mu$ is the usual 4-vector position. The SUSY extension of
(standard) gauge-invariant (Maxwell, Yang-Mills) theories are
precisely constructed  from $W$ by considering $W^2$ and its
hermitian conjugate $\bar W^2$.

Now, as stressed in \cite{DP}, another superfield combination
enters into play if one wishes to construct \underline{general}
gauge invariant SUSY Lagrangians. In particular, one needs to
consider two  superfields $X$ and $Y$ defined as
\be X = \frac{1}{8} (D^\alpha D_\alpha W^2 + \bar D_{\dot
\alpha}\bar D^{\dot \alpha}\bar W^2) \label{13v} \ee
\be Y = -\frac{i}{16}  (D^\alpha D_\alpha W^2 - \bar D_{\dot
\alpha}\bar D^{\dot \alpha}\bar W^2) \label{14v} \ee
with covariant derivatives given by \be D_\alpha = ~ \frac
\partial {\partial \theta ^\alpha }+2i \left( \sigma ^\mu\bar
\theta \right) _\alpha \frac \partial {\partial y^\mu}, \ \ \ \ \
\bar D_{\dot\alpha} = -\frac{\partial}{\partial
\bar\theta^{\dot\alpha}} \label{10v} \ee when acting on functions
of $(y,\theta,\bar\theta)$ and \be D_{\alpha} =
\frac{\partial}{\partial \theta^\alpha}, \ \ \ \ \bar D_{\dot
\alpha } = - \frac
\partial {\partial \bar \theta ^{\dot \alpha }%
}-2i\left( \theta \sigma ^\mu\right) _{\dot \alpha } \frac
\partial {\partial y^{\dagger\mu}} \label{11v} \ee on functions of
$(y^\dagger,\theta,\bar\theta)$.
The only components of these superfields having purely bosonic
terms are

\be
\left.W^2\right|_{\theta\theta} = D^2 - \frac{1}{2} F^{\mu \nu} F_{\mu \nu}
- i\frac{1}{2}
F^{\mu \nu}\tilde F _{\mu \nu}
\ee
\be
\left.W^2\right|_{\theta\theta} = D^2 - \frac{1}{2} F^{\mu \nu} F_{\mu \nu}
+ i\frac{1}{2}
F^{\mu \nu}\tilde F _{\mu \nu}
\ee

\ba X |_{0} &=& - (
 D^2 - \frac{1}{2} F^{\mu \nu} F_{\mu \nu}
-i \lambda \dsl \bar \lambda -i \bar \lambda \bar{\dsl}  \lambda)
\nonumber\\
X |_{\theta\bar\theta} &=& i\theta\sigma^p\bar\theta\partial_p\,
(D^2 - \frac{1}{2} F^{\mu \nu} F_{\mu \nu} -i \lambda \dsl \bar
\lambda -i \bar \lambda \bar{\dsl}  \lambda)
\nonumber\\
X |_{\theta\bar\theta\theta\bar\theta} &=& \frac 1 4
\theta\bar\theta\theta\bar\theta\,\Box\, (D^2 - \frac{1}{2} F^{\mu
\nu} F_{\mu \nu} -i \lambda \dsl \bar \lambda -i \bar \lambda
\bar{\dsl} \lambda) \label{3} \ea
and \ba Y|_{0} &=& \frac{1}{2} (\frac{1}{2} F^{\mu \nu}\tilde F
_{\mu \nu} +  \lambda \dsl \bar \lambda
- \bar \lambda \bar{\dsl}  \lambda)\nonumber\\
Y|_{\theta\bar \theta} &=& -\frac{i}{2}\theta\sigma^p\bar\theta
\partial_p\, (\frac{1}{2}
F^{\mu \nu}\tilde F _{\mu \nu} +  \lambda \dsl \bar \lambda
- \bar \lambda \bar{\dsl}  \lambda)\nonumber\\
Y|_{\theta\bar \theta\theta\bar \theta} &=& \frac 1 8 \theta\bar
\theta\theta\bar \theta\, \Box\, (\frac{1}{2} F^{\mu \nu}\tilde F
_{\mu \nu} +  \lambda \dsl \bar \lambda - \bar \lambda \bar{\dsl}
\lambda) \label{4} \ea
with $\tilde F_{\mu \nu} = (1/2) \varepsilon_{\mu \nu \alpha
\beta} F^{\alpha \beta}$.

A third superfield combination is necessary for constructing
general gauge invariant SUSY Lagrangian. This combination is $W^2
\bar W^2$ with its highest component taking the form
\be W^2 \bar W^2\vert_{\theta \theta \bar \theta \bar \theta} =
\theta \theta \bar \theta \bar \theta \left( (D^2 - \frac{1}{2}
F_{\mu \nu}F^{\mu \nu})^2 + (\frac{1}{2}\tilde F_{\mu \nu}F^{\mu
\nu})^2 \right) \label{5} \ee

Remark that in all three cases (\ref{3})-(\ref{5}), all   dependence  on the
curvature $F_{\mu \nu}$ and the auxiliary field $D$  is through
the combination
\be t = \frac{1}{\beta^2} \left(D^2 - \frac{1}{2} F^{\mu \nu}
F_{\mu \nu}\right) \label{6} \ee
and this fact will have important consequences in our discussion.
Here, in order to define a dimensionless variable $t$ we have
introduced a parameter $\beta$ with the same dimensions as
$F_{\mu\nu}$ (i.e. dimensions of a mass in $d=4$). It corresponds
to the {\it absolute field} in the Born-Infeld theory
\cite{B}-\cite{BI} as will become clear below.

As it happens for the last component of $W^2\bar W^2$, also the
term $W^2$ ($\bar W^2$)
 depends on $F_{\mu \nu}$ and
$D$ through the combination (\ref{6}). Indeed, the last component
in $W^2$ ($\bar W^2$) contains the term $D^2 - \frac{1}{2} F^{\mu
\nu} F_{\mu \nu} + i F_{\mu \nu}\tilde F^{\mu \nu}$ ($D^2 -
\frac{1}{2} F^{\mu \nu} F_{\mu \nu} -
 i F_{\mu \nu} \tilde F^{\mu \nu}$) so that the sum of $\theta$
 ($\bar \theta$ )integrals
leads to the well-known SUSY extension of the Maxwell theory.

We are  ready to write a general $N=1$ supersymmetric Lagrangian
endowed with gauge-invariance in terms of $X$, $Y$ and $W^2\bar
W^2$
\begin{equation}
 L^{d=4} =
\frac{1}{4e^2} \int\left( W^2 d^2\theta  + \bar W^2  d^2\bar
\theta \right) + \frac{1}{e^2}\!\!\!\sum_{r,s,t=0}^\infty \!\!\!
a_{rst}\int\!\!
d^4\theta \left( W^2\bar W^2\right)^r X^s Y^t \label{7}
\end{equation}
with $e$ the fundamental gauge coupling constant, which has been
factorized in both terms for later convenience.

 The
second term accounts for the non-polynomial features of the
general bosonic theory to be supersymmetrized. As explained in
\cite{DP}, supersymmetry imposes two constraints on coefficients
$a_{rst}$. Their explicit form  will not be relevant for our
discussion. What one should retain is that expression (\ref{7})
gives then the most general Lagrangian corresponding to the
supersymmetric extension of a general bosonic Lagrangian depending
on the two algebraic Maxwell invariants $ F^{\mu \nu} F_{\mu \nu}$
and $ \tilde F^{\mu \nu} F_{\mu \nu}$.

We shall focus now on a $d=3$, $N=2$
supersymmetric theory which can be obtained from Lagrangian
(\ref{7}) by dimensional reduction. The standard procedure for
dimensional reduction, say in the $x_3$ spatial coordinate,
implies identifying $A_3$ with a scalar field $N$. It can be
shown that without including a Chern-Simons term, the bosonic part
of the Lagrangian (\ref{7}) can only yield electrically neutral
configurations, so that as long as one looks for self-dual
equations associated with (static) vortices,  the $A_0$ field (as
well as the $N$ field) can be put to zero  and so we will do from
here on (the case $N \ne 0$ can be equally treated without
additional complications). So far,  without the addition of a
Chern-Simons term, no electrically charged vortices exist and then
the most general gauge field configurations are pure magnetic
Nielsen-Olesen type soliton solutions. This implies
that no $d=3$ version of the $\tilde F_{\mu \nu} F^{\mu \nu}$
functional are available and that we can simply identify the field
strength with the magnetic field $B$  by
\be \frac{1}{2} F_{\mu \nu} F^{\mu \nu} = B^2 \label{8} \ee
with
\be B = \frac{1}{2} \varepsilon_{jk}F^{jk}  ~ ~ ~  i,j=1,2
\label{9} \ee
Once the dimensional reduction is carried on, one ends with the
$d=3$ version of the SUSY Lagrangian given in eq.(\ref{7}). As it
is well known, supersymmetry can be extended from $N=1$ to $N=2$
in this process.

From what we have seen, the gauge field  dependent terms in the
$bosonic$ part of this $N=2$ supersymmetric Lagrangian can be
compactly written in the form
\be L_{A}[A_\mu,D] = \frac{1}{e^2}\sum_{n=0}^{\infty} c_n t^n
\label{10} \ee where $t$ (defined in (\ref{6})) now reads
\be t = \frac{D^2 - B^2}{\beta^2} \label{11} \ee
and $c_n$ are some coefficients which can be computed in terms of
the $a_{rst}$'s.

Concerning the Higgs field sector, in $d=4$ dimensions the
coupling between the scalar Higgs field $\phi$ and the gauge field
$A_\mu$ arises from the superfield interaction term
\be L_{A-\phi}^{d=4} = \Phi \exp(V) \Phi^* \label{12} \ee
where $\Phi$ is a chiral scalar superfield containing a Higgs
field $\phi$, a higgsino $\psi$ and an auxiliary field $F$.  One
can easily see that the part of $L_{A-\phi}$ containing the
auxiliary field $D$ is  \cite{ENS}
\be L_{A-\phi}^{d=4}|_D = \frac{1}{2}  D \vert \phi \vert^2
\label{13} \ee On the other hand, gauge symmetry breaking  can be
achieved {\it \`a la} Fayet-Iliopoulos  so that the complete $D$
dependence of the supersymmetric Lagrangian arising from the Higgs
coupling to $A_\mu$ and $D$ is given by
\be L^{d=4}_D [A,\phi,D] \equiv L^{d=4}_{A-\phi}|_D + L^{d=4}_{FI}
=
  \frac{1}{2} D(|\phi|^2 - \xi^2)
\label{14} \ee
where $\xi$ is a real constant.
This Lagrangian remains unchanged after dimensional reduction so
that  we can write  the  $D$ dependent terms of the $d=3$  bosonic
part of the Lagrangian as
\be L_D^{total}[A,\phi,D] = \frac{1}{e^2} \sum_{n=0}^{\infty} c_n
\left(\frac{1}{\beta^2}( D^2 - B^2)\right)\!^n  +
 \frac{1}{2} D(|\phi|^2 - \xi^2)
\label{15} \ee
In $d=3$ space-time, dimensions of parameters and fields are
$[\beta]= m^2$, $[e] = m^{\frac{1}{2}}$, $[\xi] =
m^{\frac{1}{2}}$, $[A_\mu] = m$, $[D] = m^2$ and $[\phi] =
m^{\frac{1}{2}}$. Then, for dimensional reasons, one can infer
that coefficients $c_n's$ can be written in the form
\be c_n = \beta^2 \lambda_n \label{17} \ee
where $\lambda_n$ are dimensionless coefficients.

We can now obtain the equation of motion for $D$ so as to
eliminate the auxiliary field from the physical spectrum
\be \sum_{n=0}^{\infty}  \frac{2n}{e^2} \lambda_n
\left(\frac{1}{\beta^2}( D^2 - B^2)\right)\!^{n - 1} {D} +
 \frac{1}{2}(|\phi|^2 - \xi^2) = 0
\label{16} \ee
One can easily see that the only nontrivial solution to
eq.(\ref{16}) takes the form
\begin{eqnarray}
D & = & - \frac{e^2}{4\lambda_1}(|\phi|^2 - \xi^2 \nonumber)\\
B & = & \pm D \label{17v}
\end{eqnarray}
These two equations can be readily combined into one  which is
nothing but the well-honored Bogomol'nyi equation for the magnetic
field of the Nielsen-Olesen vortices
\be B = \mp \frac{e^2}{4\lambda_1}(|\phi|^2 - \xi^2) \label{18}
\ee
This shows that the Bogomol'nyi gauge field equation for vortex
configurations is \underline{independent} of the particular form
of the gauge field Lagrangian one chooses since we have proven
formula (\ref{18}) for the general supersymmetric Lagrangian
(\ref{7})+(\ref{12}). It should be
noted that the choice of $\lambda_1 = -1/2$, $\lambda_n = 0$ for
$n \ne 1$ corresponds to the usual value of the Maxwell term while
the choice $\lambda_1 = -1/2$, $\lambda_2 = 1/8$, $\lambda_3  =
1/32$, $\ldots$, gives a Dirac-Born-Infeld Lagrangian for the gauge
field, which, once the auxiliary fields are put on shell takes the form
\be
L_{DBI} = - \beta^2\left(\sqrt{ 1 + \frac{1}{2\beta^2}F_{\mu\nu}F^{\mu\nu}}
-1 \right)
\ee

Let us now analyze the $N=2$ supersymmetry transformations
leaving invariant the three dimensional
Born-Infeld SUSY theory. 
We shall not write the complete set of transformations but just
those which are relevant for the discussion of Bogomol'nyi
equations, namely those for the higgsino and gaugino (which we
call $\psi$ and $\Sigma$):
\be \delta_\epsilon \psi = -i \Dsl \phi \epsilon
= \left( \begin{array}{cc} 0 & D_1 + i D_2\\
                       D_1 - i D_2 & 0 \end{array} \right)
\left( \begin{array}{c} \epsilon_+\\
                        \epsilon_- \end{array} \right)
\label{19} \ee
\be \delta_\epsilon \Sigma =
(\frac{1}{2}\varepsilon_{\mu\nu\alpha} F^{\mu\nu}\gamma^\alpha +
D) \epsilon =
 \left( \begin{array}{cc} \frac{1}{2} \varepsilon_{ij}F^{ij} + D & 0\\
                      0 & \frac{1}{2} \varepsilon_{ij}F^{ij} - D
\end{array} \right)
\left( \begin{array}{c} \epsilon_+\\
                        \epsilon_- \end{array} \right)\\
\label{20} \ee
where we call $\epsilon$ the Dirac fermion transformation
parameter (we have already made $N = A_0 = 0$ and considered the
static case).

As it is well-known by now, making zero half of the SUSY variations
associated with the higgsino and gaugino fields, one gets the
Bogomol'nyi equations. For instance, by demanding that those
generated by $\varepsilon_+$ be zero, one gets the following
 self-dual equation from the higgsino's
variation
\be \delta_{\epsilon_+} \psi = 0 \to D_1 \phi = i D_2 \phi
\label{21} \ee
One should note that this transformation law just depends on the
way the parallel displacement is defined in terms of the gauge
connection and not on the explicit form of the gauge field action.
One can then understand why eq.(\ref{21}) is completely
independent of the particular form the gauge field action is
chosen, at least for minimally coupled gauge theories
\footnote{For an analysis of Bogomol'nyi equations in
non-minimally coupled gauge theories, see ref.\cite{cbpf}.}.
Regarding the equation derived from the gaugino transformation,
\be \delta_{\epsilon_+} \Sigma = \frac{1}{2}
\varepsilon_{ij}F^{ij} + D = 0 \label{22} \ee
it could, in principle, depend on the particular Lagrangian chosen
through the $D$ term. However, as we have seen (eq.(\ref{17v})),
the solution to the  equation of motion for $D$ takes the same
simple form for any gauge field Lagrangian since $D$ always enters
through the combination $D^2 - B^2$.

This feature can be also checked by analyzing the two supercharges
which can be obtained following the usual Noether construction. As
it has been shown in \cite{GNSS} for the Born-Infeld case,
supercharges $Q$ and $\bar Q$ can be always put in the form
\ba \bar Q &=& i \int d^2x\ \Sigma^\dagger\ {\cal H}[B,D]\
(\gamma^0 B +D) +
\frac{i}{2} \int d^2x\ \psi^\dagger\ {\Dsl \phi}\nonumber \\
Q &=& -i\int d^2x\  (B +\gamma^0 D)\ {\cal H}[B,D]\ \Sigma -
\frac{i}{2} \int d^2x\ \gamma^0 ({\Dsl \phi})^\dagger \psi
\label{24} \ea
with ${\cal H}$ some real functional of $D$ and $B$ which can be
computed order by order in $1/\beta^2$. Furthermore,
eqs.(\ref{24}) also hold when one considers not just SUSY
Born-Infeld theory but the general Lagrangian, viz. eq.(\ref{7}).
Only the actual form of ${\cal H}$ will change, depending on the
different sets of possible $a's$ coefficients. What one can easily
see is that the following formula holds
\be {\cal H} = {\cal H}_{Maxwell}+
\sum_{n=1}^{\infty}\frac{1}{\beta^{2n}} {\cal H}_n[B,D]
\label{gene} \ee
with
\be {\cal H}_{Maxwell} = 1 \label{mu} \ee
\be \left.{\cal H}_n [B,D]\right\vert_{B^2 = D^2} = 0 \label{que}
\ee

It is clear that condition $\bar Q |phys\rangle = 0$ is satisfied
whenever $(B  +\gamma^0 D)\epsilon =0$ and  $ {\Dsl\phi} \epsilon
= 0$, independently of the precise form the functional ${\cal H}$
takes. Choosing just the upper component of the transformation
parameter, $\epsilon_+$, yields again the two Bogomol'nyi
equations (\ref{18}),(\ref{21}). Of course, this is to be expected
since both $(B  + \gamma^0 D)\epsilon$ and $ \Dsl \phi \epsilon$,
appearing in (\ref{24}), provide the transformation laws of
gaugino and higgsino respectively.

Concerning the supercharge algebra,  when the Bogomol'nyi equation
$B = \pm D$ is used, only the Maxwell part of ${\cal H}$ survives,
this showing again why the BPS structure is not sensitive to the
particular form of the gauge field Lagrangian.

 ~

In conclusion, we have analyzed the most general Lagrangian
corresponding to the supersymmetric extension of a general bosonic
Lagrangian depending on the two algebraic Maxwell invariants $
F^{\mu \nu} F_{\mu \nu}$ and $ \tilde F^{\mu \nu} F_{\mu \nu}$.
This general Lagrangian includes, for a particular choice of
coefficients, the Born-Infeld supersymmetric Lagrangian, and also
an infinite class of Lagrangians having causal propagation
\cite{DP}. We have shown why the Bogomol'nyi relations associated
with the bosonic sector remain unchanged in spite of the actual
form of the gauge field Lagrangian: Maxwell, Born-Infeld or more
complicated non-polynomial Lagrangians all have the same BPS
structure.

\section{Including axions}

Cosmic strings appearing as topological defects in Grand Unified Theory
aroused a lot of interest about twenty years ago
as a source of primordial density perturbations from which galaxies eventually
grew. In this scenario, since Abrikosov-Nielsen-Olesen strings arise
in any gauge theory with a broken $U(1)$ symmetry, such strings
could appear whenever a $U(1)$ symmetry becomes broken as  the Universe
cools.

However, after the microwave background results from COBE,
BOOMERanG and WMAP  it was accepted, at the end of the 90's,
that cosmic strings or other topological defects arising at GUT scales
could not provide an explanation for the bulk of the density perturbations.

The possibility of microscopic fundamental strings in superstring theories
acting as seeds for galaxy formation was also excluded in the context
of perturbative string theory \cite{witt}: among other problems,
their tension $\mu$ ,
close to Planck scale, would produce inhomogeneities in the cosmic microwave
background far larger
than observed. For example in perturbative heterotic string theory
$G\mu = \alpha_{GUT}/16\pi \geq 10^{-3}$ while the isotropy of the cosmic
microwave background implied $G\mu = \alpha_{GUT}/16\pi
\leq 10^{-3}$ \cite{HK}

The question was be recently reconsidered,  after the relevance of
branes and new kind of extended objects was understood (see \cite{pol}
and references therein) opening the possibility to the string tension
 to be much lower, anything between the Planck scale and the weak scale.
In order to understand the nature and structure
of such {\it stringy} cosmic strings, the embedding  of BPS objects
in supersymmetry and supergravity models has become an active area of
research so that the properties of BPS solitons,
their connections with the supersymmetry algebra and their cosmological
 applications have been  discussed by many authors \cite{ENS},\cite{EdelNS}-\cite{Gorsky}.

Having in mind the study of BPS solitons in a string theory context,
where the axion is almost unavoidable, it is natural to consider
gauge-Higgs systems in which the axion field is included. In particular,
an  $N=1$ globally
supersymmetric model in $d=3+1$ dimensions consisting of an axion
superfield $S$
 coupled to $W_\alpha W^\alpha$, with $W_\alpha$ the chiral  superfield
 strength, was analyzed in  \cite{Blanco} and finite energy cosmic string solutions
  for the resulting bosonic Maxwell-Higgs action coupled to an
   axion field were constructed. Also, the impact of axions on dynamics of
   a $d=3+1$ Yang-Mills theory supporting non-Abelian strings has been analyzed
in \cite{Gorsky}.

We shall consider here this issue taking as an example a
 $d= 2+1$ space-time dimensions for which a rich variety of flux
tube solutions exists already when the axion field is absent.
Indeed, as we have seen, when gauge fields with dynamics governed by a Chern-Simons
action are coupled to charged scalars with an appropriate sixth
order
  symmetry breaking potential,
 the model
admits BPS equations with vortex-like solutions carrying both
magnetic flux and electric charge \cite{Hong}-\cite{JW}. It should
be stressed that in the absence of the Chern Simons term,
electrically charged vortices with finite energy (per unit length)
do not exist \cite{JZ}. Hence, the model we are interested in
could show novel aspects of charged string like configurations
when an axion is present, in particular with respect to their
application to cosmological problems. Of course, this in the
perspective that at high temperatures, a relativistic four
dimensional quantum field theory becomes effectively three
dimensional.

\subsection*{The model}

The coupling of an axion to a gauge field with dynamics governed by a
Chern-Simons action poses some problems \cite{Burgess}-\cite{FG}.
To discuss
how can they can be overcome, let us consider the following
$(2+1)$-dimensional bosonic action,

\begin{equation}
{\cal S} = \int d^3x\, \left\{\frac{\kappa}{8\delta}f(s){\rm
Im}D_\mu S\tilde F^\mu + \vert D_\mu \phi\vert^2 +
K^{\prime\prime}(s) \vert D_\mu S\vert^2 - W(\phi, S)\right\}
\label{accion}
\end{equation}
where $\phi$ and $S = s + ia$ are complex fields, $A_\mu$ is a
$U(1)$ gauge field, and $\tilde F^\mu$ is defined as,
\begin{equation}
\tilde F^\mu \equiv \epsilon^{\mu\nu\sigma}\partial_\nu A_\sigma
\end{equation}
$W(\phi,\phi^*,S,S^*)$ is a potential term, $f$ and $K$ are
arbitrary functions of $s$, the real part of $S$, primes stand for
$\partial/\partial s$, $\kappa$ is a constant and $\delta$ is a
dimensionless parameter (which in the
4 dimensional case is related to the Fayet-Iliopoulos term and the Planck
mass). Finally, $D_\mu$ is the covariant derivative acting on the
fields $\phi$ and $S$ according to
\begin{eqnarray}
D_\mu \phi &=& \partial_\mu \phi - ieA_\mu\phi\nonumber\\
D_\mu S &=&  \partial_\mu S + 2i\delta A_\mu
\end{eqnarray}
As done in \cite{Blanco} for the $3+1$ model,
 we shall identify $a$ in (\ref{accion}) with the axion field
and $s$ with a dilaton field. Note that because of the definition of the
axion covariant
 derivative,  the Chern-Simons
term appears in action  (\ref{accion}) multiplied by the factor $f(s)$,
 \begin{eqnarray}
{\cal S}_{CS}[A,s] = \frac{\kappa}{4} \int d^3x f(s)
 \epsilon^{\mu\nu\sigma} A_\mu \partial_\nu A_\sigma
 \label{normal}
 \end{eqnarray}

The action (\ref{accion}) is invariant under the local
transformation
\begin{eqnarray}
S ~ &\longrightarrow& S - 2i\delta \Lambda(x)\nonumber\\
S^*&\longrightarrow& S^* + 2i\delta \Lambda(x)\nonumber\\
\phi ~ &\longrightarrow & e^{ie\Lambda(x)}\phi\label{gauge}\\
\phi^*&\longrightarrow &\phi^* e^{-ie\Lambda(x)}\nonumber\\
A_\mu &\longrightarrow & A_\mu + \partial_\mu \Lambda(x)\nonumber
\end{eqnarray}

The time component of the gauge field equation of motion is the
Chern-Simons version of the Gauss law and can be used to solve for
$A_0$ giving
\begin{equation}
A_0 = \frac{\kappa\,{\cal B}}{2(e^2\vert\phi\vert^2 +
4\delta^2K^{\prime\prime})} \label{gausi}
\end{equation}
where
\begin{equation}
{\cal B} \equiv = f(s)F_{xy} - \epsilon^{ij}(A_i +
\frac{1}{2\delta}\partial_ia)\partial_jf(s)
\end{equation}
The energy can be found from the energy-momentum tensor obtained
by varying the curved space form of the action with respect to the
metric,
\begin{equation}
\delta S = \frac{1}{2} \int d^3x \sqrt g \, T^{\mu\nu} \delta
g_{\mu\nu}
\end{equation}
 Integration
of the time-time component $T^{00}$ gives
\begin{equation}
{\cal E}= \int d^2x\,\left\{ \vert D_i\phi\vert^2 +
K^{\prime\prime}\vert D_i S\vert^2 + W(\phi,S) + \frac{\kappa^2
{\cal B}^2}{4(e^2\vert\phi\vert^2 +
4\delta^2K^{\prime\prime})}\right\}
\end{equation}
After some work, this expression
can be written in the form
\begin{eqnarray}
&& \hspace{-1 cm} {\cal E} = \int \!\!d^2x\!\!\left(
\!\vphantom{\frac{\left(e^2\vert\phi\vert^2 + 4\delta^2
K^{''}\right)}{s/2}} \vert D_x\phi\pm iD_y\phi\vert^2 +
K^{\prime\prime}\vert D_xS \pm iD_y S\vert^2 +
\frac{\kappa^2}{4(e^2\vert\phi\vert^2 + 4\delta^2K^{''})}\times
\right.
\nonumber\\
&&
\left. \left[{\cal B} \pm\frac{\left(e^2\vert\phi\vert^2 +
4\delta^2 K^{''}\right)}{\kappa^2\,f(s)/2}
\left(e(\vert\phi\vert^2-\vert\phi_0\vert^2)-4\delta K^{\prime}
\right)\right]^2\right.\nonumber\\
&&\left. + W - \frac{1}{(\kappa\,f(s))^2}\left(e^2\vert\phi\vert^2
+ 4\delta^2
K^{''}\right)\left(e(\vert\phi\vert^2-\vert\phi_0\vert^2)-4\delta
K^{\prime}
\right)^2\right.\nonumber\\
&&  \pm \left(e(\vert\phi\vert^2-\vert\phi_0\vert^2)-4\delta
K^{\prime} \right)\epsilon^{ij}\left[A_i +
\frac{1}{2\delta}\partial_ia\right]\partial_j{\rm log} f(s)
  \nonumber\\
& & \left.  \vphantom{\frac{\left(e^2\vert\phi\vert^2 + 4\delta^2
K^{''}\right)}{s/2}}\pm e\vert\phi_0\vert^2F_{xy} \right)
\label{bogomolnyi}
\end{eqnarray}
We thus see that the first two lines in (\ref{bogomolnyi}) have
been accommodated as perfect squares. This, together with an
appropriate choice of the potential so as to cancel the third line,
would lead to a Bogomol'nyi bound for the energy given by the
magnetic flux $\Phi$ appearing in the last line,
\begin{equation}
\Phi = \int d^2x F_{xy}
\end{equation}
 There is however the term in the forth line in (\ref{bogomolnyi}) with
 no definite sign preventing the obtention of a bound.
 Only if we put $f(s) = 1$, which
corresponds to a normal Chern-Simons action for the gauge field
(see eq.(\ref{normal})),
this term vanishes. In that case one does have a bound,
\begin{equation}
{\cal E} \geq \pm e\vert\phi_0\vert^2 \Phi = 2\pi
e\vert\phi_0\vert^2  |n | \label{boundxx}
\end{equation}
whenever the potential is chosen as
\begin{equation}
W =
 \frac{1}{\kappa^2} \left(e^2\vert\phi\vert^2 + 4\delta^2 K^{\prime\prime} \right)
\left(e(\vert\phi\vert^2-\vert\phi_0\vert^2) - 4\delta K^{\prime}
\right)^2\label{potential}
\end{equation}
The bound is saturated by fields obeying the self-duality
equations
\begin{eqnarray}
&& D_x\phi = \mp iD_y\phi  \nonumber\\
&& D_x S = \mp iD_y S \nonumber\\
&& \kappa^2 F_{xy} = \mp 2 \left(e^2\vert\phi\vert^2 + 4\delta^2
K^{\prime\prime}  \right)
\left(e(\vert\phi\vert^2-\vert\phi_0\vert^2) - 4\delta K^\prime
\right)
\label{bpseq}
\end{eqnarray}
where the upper (lower) sign corresponds to positive (negative)
values of $\Phi$ and $n$.

As it is well-known, the presence of a Chern-Simons term forces a
relation between magnetic flux and electric charge \cite{DJT}.
This makes the Chern-Simons vortices both magnetically and
electrically charged. To see this phenomenon in the present case
let us write the gauge field equation of motion, for the case
$f(s)=1$ in the form
\begin{equation}
\frac{\kappa}{2} \varepsilon^{\mu\alpha\beta}F_{\alpha\beta} =
J^\mu \label{3ese}
\end{equation}
with $J^\mu= (\rho, \vec J)$ the conserved matter current and
$\rho$ the electric charge density,
\begin{eqnarray}
  J^\mu = - 2\left(e^2 |\phi|^2 + 4K''\delta^2\right)A^\mu -
  ie(\phi\partial^\mu\phi^* + \phi^*\partial^\mu\phi) -
 4\delta
  \partial^\mu a
\end{eqnarray}
We then see that eq.(\ref{gausi}) can be rewritten in the form
\begin{equation}
\rho = -\kappa {\cal B}
\end{equation}
so that the usual relation between electric charge $Q = \int d^2x
\rho$  and magnetic flux in Chern-Simons theories holds,
\begin{equation}
\label{related}
Q = -\kappa \Phi
\end{equation}
Note that both the Higgs scalar and the axion contribute to the
electric charge.

\section*{Supersymmetric extension}

The SUSY extension of the Chern-Simons-Higgs system with a sixth order
symmetry breaking potential was analyzed
in \cite{LLW}. Let us study now the case in which the axion
field is also present.

To do  so, we will consider the following $d=3$ action, written in
terms of superfields as
\begin{eqnarray}
{\cal S}_{{\rm SUSY}} &=&-\frac{1}{2}\int d^3x\,d^2\theta \left(\frac{\kappa}{4\delta}
F(\Sigma+\Sigma^\dag){\rm Im}\left(\tilde\nabla_a\Sigma\right)W^a
\right.\hspace*{2cm}\nonumber\\
&+&\left. H(\Phi^\dag,\Phi)(\nabla^a\Phi)^\dag\nabla_a\Phi+ K_{s\bar s}(\Sigma,\Sigma^\dag)\vert \tilde\nabla_a\Sigma\vert^2 \right.\nonumber\\
&+&\left.2V(\Phi^\dag,\Phi, \Sigma,\Sigma^\dag)
\right)\label{suaction}
\end{eqnarray}
Here  $\mu=0,1,2$ and $a=1,2$. $\Gamma_a$, $\Phi$ and $\Sigma$ are spinor, complex scalar
and axionic superfields and
$$\nabla_a \Phi = (D_a - ie \Gamma_a)\Phi $$
$$\tilde \nabla_a \Sigma = D_a\Sigma + 2i\delta \Gamma_a $$
$$D_a = \frac{\delta}{\delta\theta^a} + i (\gamma^\mu)_{ab}\theta^b\partial_\mu$$
The lowest component in $\Gamma^a$ is the gauge field, that in $\Phi$ corresponds to the Higgs field and that in $\Sigma$ is $S=s + i a$ with $a$ the axion field.
Concerning $F, H, K$ and $V$, they are functionals of
superfields to be fixed later. Subindexes in these functionals
mean derivatives,
thus $K_{s\bar s} = \partial_S\partial_{S^*}K=
\partial_\Sigma\partial_{\bar\Sigma} K\vert_{\theta=0}$ and so on.
They should be chosen so that the
supersymmetric action (\ref{suaction}) is invariant under
the supergauge transformations,
\begin{eqnarray}
\Phi&\longrightarrow& e^{i e \Lambda}\Phi\nonumber\\
\Gamma_a&\longrightarrow &\Gamma_a - D_a \Lambda \nonumber\\
\Sigma &\longrightarrow &\Sigma_a  - 2\delta \Lambda
\label{susytrafo}
\end{eqnarray}
for any real scalar superfield $\Lambda$.
Written in the Wess-Zumino gauge, {\it i.e.},
$$\Gamma_a\vert_{\theta=0}= D^a\Gamma_a\vert_{\theta =0}= 0,$$
the spinor superfield $\Gamma_a$
is given by
$$\Gamma_a(x,\theta) = i\theta^b(\gamma^\mu)_{ba} A_\mu(x) - 2
\theta^2 \lambda_a(x)$$
where $\lambda(x)$ is a Majorana spinor, the
photino. Then, the spinor field strength, defined as
$$W_a = \frac{1}{2} D^bD_a \Gamma_b$$
  takes, in terms of component fields, the form
$$W_a(x,\theta) = \lambda_a(x) - \frac{1}{2}
\theta^b(\gamma^\mu\gamma^\nu)_{ba}F_{\mu\nu} - i
\theta^2(\gamma^\mu)^b_a \partial_\mu\lambda_b(x)$$
and satisfies the Bianchi identity,
$$D^aW_a = 0.$$
The  complex scalar superfield is defined as
$$\Phi(x,\theta) = \phi(x)+ \theta^a\psi_a(x) - \theta^2 F(x)
$$
where $\phi$ stands for the Higgs complex scalar field, $\psi_a$
is a Dirac bispinor, the higgsino and $F$ is a
complex auxiliary field. Finally, the superfield $\Sigma$ which
contains the axion $S$ as its lowest component,
can be reduced to a complex scalar field
$\Theta$ by exponentiation,
$$ \Theta = e^{\Sigma}$$
The supersymmetric transformations
(\ref{susytrafo}) take $\Gamma^a$
out from the Wess-Zumino gauge. One can however
implement a composition of SUSY and gauge
transformations such that the Wess-Zumino gauge
remains valid. To
do that, the new SUSY transformation for scalar
and spinorial
superfields are, respectively,
\begin{eqnarray}
\delta^{\rm WZ}_\eta \Gamma_a &=& i\eta^b Q_b
\Gamma_a +
D_a\tilde K\nonumber\\
\delta^{\rm WZ}_\eta \Phi &=& i\eta^b Q_b \Phi +
i\tilde K\Phi\label{transf}\\
\delta^{\rm WZ}_\eta \Sigma &=& i\eta^bQ_b \Sigma
+ i\tilde K \nonumber
\end{eqnarray}
where $Q_a=i\partial_a+\theta^b(\gamma^\mu)_{ba}
\partial_\mu$, $\eta^a$ a Majorana spinor and the real scalar
superfield $\tilde K$ is defined as,
$$\tilde K = i\theta^a(\gamma^\mu)_{ab}\eta^b A_\mu + \theta^2\lambda^a\eta_a.$$
Let us restrict the action (\ref{suaction}) to the
case in which
\begin{equation}
F(\Sigma+\Sigma^\dag) = H(\Phi^\dag,\Phi) = 1
\end{equation}
Then, written in components, the action takes the simple form
\begin{equation}
{\cal S}_{SUSY} = {\cal S}_{\rm B} + {\cal S}_{\rm
F}
\end{equation}
with the pure bosonic action given by
\begin{equation}
{\cal S}_{\rm B}=\int d^3x\,\left\{ \frac{\kappa}
{4}\epsilon^{\mu\nu\sigma}A_\mu F_{\nu\sigma} +
\vert D_\mu\phi\vert^2 + R\vert\tilde D_\mu
S\vert^2 -\vert V_\phi\vert^2 -\frac{1}{R}\vert
V_s\vert^2\right\}
\label{boson}
\end{equation}
while  the fermionic one is given by
\begin{eqnarray}
{\cal S}_{\rm F}&=&\int d^3x\,\left\{
i\bar\psi\gamma^\mu{\mathop{D}^\leftrightarrow}
_\mu\psi + i R\bar\chi\gamma^\mu
{\mathop{\partial}^\leftrightarrow}_\mu\chi
-\frac{1}{2} R_{s\bar s}
(\bar\chi\chi)^2\right.\nonumber\\
&+&
\frac{1}{2}
\left[i\bar\chi\gamma^\mu\chi(R_s\,\tilde D_\mu S
- R_{\bar s}\,\tilde D_\mu S^*)\right]\nonumber\\
&+&\bar\psi\psi\left(\frac{e^2}{\kappa}
\vert\phi\vert^2
- V_{\phi\bar\phi}\right)
+\bar\chi\chi\left(\frac{4\delta^2}{\kappa}
R^2 - V_{s\bar s}\right)
\nonumber\\
&-& \bar\psi\chi\left(\frac{2\delta e}{\kappa}\phi
R + V_{\bar\phi s}\right)-
\bar\psi^*\chi^*\left(\frac{2\delta e}{\kappa}
\phi^*
R + V_{\phi\bar s}\right)- \frac{\vert R_s\vert^2}
{4R}\vert\bar\chi\chi^*\vert^2\nonumber\\
&+& \frac{1}{2}\bar\psi^*\psi\left(
- V_{\phi\phi} -\frac{e^2}{\kappa}\phi^{*2}
\right)+ \frac{1}{2}\bar\psi\psi^*\left(
- V_{\bar\phi\bar\phi}
-\frac{e^2}{\kappa}\phi^{2}
\right)
\nonumber\\
&+& \frac{1}{2}\bar\chi^*\chi\!\!\left(\!\frac{R_s}{R}\,
V_s - V_{ss} -\frac{4\delta^2}{\kappa}
R^2\!\right)\!+\! \frac{1}{2}
\bar\chi\chi^*\!\left(\!\frac{R_{\bar s}}{R}
\,V_{\bar s} - V_{\bar s\bar s}
-\frac{4\delta^2}{\kappa}R^2\!\right)
\nonumber\\
&+& \left.\bar\psi^*\chi\left(\frac{2\delta e}
{\kappa}
\phi^*
R - V_{\phi s}\right)
+\bar\psi\chi^*\left(\frac{2\delta e}{\kappa}
\phi
R - V_{\bar\phi\bar s}\right) \right\},
\label{fermion}
\end{eqnarray}
where we have redefined $R\equiv K_{s\bar s}$.
Here we have taken into account that
\[
I =
\int d^3x d^2\theta\,\left( D^a\Sigma\right) W_a
\]
is a surface term which does not modify the
equations of motion
\begin{eqnarray}
I &=& \int d^3x d^2\theta\,
D^a\left(\Sigma W_a\right) = \int d^3x\,
D^2D^a\left.\left(\Sigma
W_a\right)\right\vert_{\theta=0}\nonumber\\
&=& = i\int
d^3x\,\partial_b^aD^b\left.\left( \Sigma
W_a\right)\right\vert_{\theta=0}\,,
\nonumber\end{eqnarray}
and can hence be neglected.

 In order to extend the $N=1$
supersymmetry to an $N=2$ supersymmetry, we shall follow \cite{LLW}
and allow the transformation parameter $\eta$
to be complex, {\it i.e.}, $\eta^*\not=\eta$.
Now, this is the same as making an $N=1$ SUSY
transformation followed by a $U(1)$ fermion phase
rotation. Thus, the new transformation for
fermions will be
$\psi_a
\to e^{i\alpha}\psi_a $ and
$\psi^*_a \to e^{-i\alpha}
\psi^*_a $ and the same for $\chi_a$ and $\chi^*_a$. The new SUSY
transformations act then as rotations on the
fermions and one can then see that the only terms
which do not respect the extended SUSY invariance are
those on the last three lines in
(\ref{fermion}). Hence, in order to get an $N=2$
supersymmetric model we need that,
\begin{eqnarray}
- V_{\phi\phi} -\frac{e^2}{\kappa}\phi^{*2}
= 0 &\hspace{.7cm},\hspace{.7cm}&\frac{2\delta e}
{\kappa}
\phi^*
R - V_{\phi s}= 0\nonumber\\
\frac{4\delta^2}{\kappa}
R^2 - V_{s\bar s } -\frac{4\delta^2}{\kappa} R^2=
0&\hspace{.7cm},\hspace{.7cm}&\frac{2\delta e}
{\kappa}
\phi
R - V_{\bar\phi\bar s} = 0\nonumber
\end{eqnarray}
where $V=V(u,v)\equiv V(\phi^*\phi, S + S^*)$.
These equations imply that,
$$ V_u = -\frac{e}{\kappa}\left(eh(u) - 2\delta
r(v)\right) H (u)$$
$$ V_v = -\frac{2\delta}{\kappa}\left(2\delta
r(v) - e h(u)\right) R(v)$$
where
\begin{equation}
\frac{d}{dv}r(v) = R(v) \; , \;\;\; \frac{d}
{du}h(u) = 1.
\end{equation}
We obviously have $h=u-u_0 = \vert\phi\vert^2
-\vert\phi_0\vert^2$, and since
$S+S^*=2s$, then $R = K^{\prime\prime}$ and $r = 2
K^\prime$ where primes stand for derivatives with
respect to $s$. From (\ref{boson}) the
potential is,
$$W =
\vert V_\phi\vert^2 + \frac{1}
{R}\vert V_s\vert^2 = \frac{1}{\kappa^2}
(e^2\vert \phi\vert^2 +
4\delta^2K^{\prime\prime})
\left[e(\vert\phi\vert^2 - \vert\phi_0\vert^2)
- 4\delta K^\prime\right]^2$$
which is exactly what we obtained in
(\ref{potential}).\par
\noindent In order to get the Bogomol'nyi bound and the self-dual equations
one can analyze  the supercharge algebra as in \cite{EdelNS}.
Alternatively, one can directly consider
the component field SUSY transformations    $(\delta_\eta X = \eta^a\delta_a X$),
\begin{equation}
\begin{array}{l c l l c l}
\delta_a A_\mu&=&-i (\gamma_\mu)_a^b \lambda_b,
&\delta_a\lambda_b &=& \frac{1}{2}
(\gamma^\mu\gamma^\nu)_{ab}F_{\mu\nu}\\
\delta_a\phi&=& - \psi_a,&\delta_a S &=&-
\chi_a,\\
\delta_a\psi_b&=& \epsilon_{ab} F - i
(\gamma^\mu)_{ab} D_\mu\phi,&
\delta_a\chi_b &=&\epsilon_{ab} J -
i(\gamma^\mu)_{ab}\tilde D_\mu S,\\
\delta_a F&=& i (\gamma^\mu)_a^b D_\mu\psi_b +
2\lambda_a\phi,&
\delta_a J &=&i (\gamma^\mu)_a^b\partial_\mu\chi_b
+ 2\lambda_a,
\label{doblecol}
\end{array}
\end{equation}
and their complex conjugated ($\delta X^\dag =
\eta^{*a}\delta_a X^\dag$) and
reobtain
Bogomol'nyi equations  (\ref{bpseq}) just by
putting all fermion fields to zero in (\ref{doblecol})
and then ask the SUSY transformations for $\psi_a$ and
$\chi_a$ to vanish once the auxiliary fields have been
written in terms of dynamical fields using their equations of motion. The first
condition corresponds to a restriction to the bosonic sector, the second
one implies that physical states are supersymmetry invariant.

\section*{String-like solutions}

We present in this section some vortex solutions to the BPS
equations of motion. We choose for the Kh\"aler potential the form
\begin{equation} K=-M^2\log\left(S+S^*\right) \end{equation}

As in \cite{Blanco}, we shall analyze separately two cases: first,
we consider the case in which the vortex is supported by the Higgs
field (``$\phi$-strings'' solutions) , in the sense that
at infinity it behaves as in the ordinary
Nielsen-Olesen vortex, with its winding number linked to
the magnetic flux. Then, we shall consider the case in which the vortex is
supported by the axion field, a solution that we shall call  an ``s-string''.
In this case it is the axion winding number which is related to the magnetic
flux.

In the first case, in order to obtain  $\phi$-string
solutions we make the ansatz \cite{Blanco}
\begin{equation}
\phi_1=f(r)e^{in\theta} \ \ \ \ \ \ \ \ \ S=s(r)-2i\delta m\theta
\ \ \ \ \ \ \ \ \ A_{\theta}=n\frac{v(r)}{r} \label{ansatz}
\end{equation}
where $n$ is the topological charge of the Higgs and $m$ is the
topological charge of the axion.

It is convenient to work with dimensionless variables by defining
\begin{eqnarray}
&&\tau=\alpha r,\ \ \ \ \ \ \ \alpha=\frac{e^2\phi_0^2}{\kappa} \nonumber\\
&&x(\tau)=v(\tau/\alpha) \nonumber\\
&&y(\tau)=e\delta^{-1} s(\tau/\alpha) \nonumber\\
&&z(\tau)=\phi_0^{-1} f(\tau/\alpha)
\end{eqnarray}
With this convention the equations read,
\begin{eqnarray}
x'&=&-\frac{2\tau}{\left|n\right|} \left(z^2 +
\frac{4\beta}{y^2}\right) \left(z^2 - 1 + \frac{4\beta}{y}\right)
\label{eqx}\\
y'&=&-\frac{2}{\tau}(\left|m\right|-\left|n\right|x) \label{eqy}\\
z'&=&\frac{z}{\tau}(1-x)\left|n\right| \label{eqz}
\end{eqnarray}
where $\beta=M^2/\phi_0^2$. From the first two equations we can
integrate $y(\tau)$ in terms of $z(\tau)$, obtaining
\begin{equation}
y(\tau)=2(\left|n\right|-\left|m\right|)\log\tau-2\log z(\tau)+k
\end{equation}
with $k$ an arbitrary integration constant. Thus, we end with a
system of two first-order coupled differential equations for
$x'(\tau)$ and $z'(\tau)$.

The boundary conditions for equations (\ref{eqx}) and (\ref{eqz})
can be determined as follows. The function $x$ must vanish at the
origin, so eq.(\ref{eqz})implies that $z$ also vanishes at the
origin as $\tau^{|n|}$. For large $\tau$, the function $x$ tends
to $1$, thus $z$ also tends to $1$ unless $|n|=|m|$. In this last
case, $z\, \to \, z_0$, where $z_0$ is the solution of the
algebraic equation
\be z_0^2 + \frac{4 \beta}{k-2 \log z_0 } =1 \ee

To solve the differential equations we employ a relaxation method
for boundary values problem. Such a method determines the solution
by starting with an initial guess and improving it iteratively.

Already from ansatz (\ref{ansatz}) one can see that the magnetic
flux and electric charge ar quantized according to
\begin{equation}
\Phi = \frac{2\pi}{e} n   \; , \;\;\; Q =-\frac{2\pi\kappa}{e} n
\end{equation}
There is an interesting property, typical of Chern Simons vortices
that also holds in our model: both the magnetic field and the
(radial) electric
field are concentrated in rings surrounding the zeroes of the Higgs field.

We also solved the BPS equations when the axion field $s$
tends asymptotically to a constants (``s-string", \cite{Blanco}).
In this case the magnetic charge is equal to the topological
charge of the axion field, so we have as an appropriate ansatz,
\begin{equation}
\phi=f(r)e^{in\theta} \ \ \ \ \ \ \ \ \ S=s(r)-2i\delta m\theta
\ \ \ \ \ \ \ \ \ A_{\theta}=m\frac{v(r)}{r}
\end{equation}
With this ansatz the BPS equations take the form
\begin{eqnarray}
x'&=&-\frac{2\tau}{\left|m\right|} \left(z^2 +
\frac{4\beta}{y^2}\right) \left(z^2 - 1 + \frac{4\beta}{y}\right)
\label{eqsx}\\
y'&=&-\frac{2}{\tau}\left|m\right|\left(1-x \right) \label{eqsy}\\
z'&=&\frac{z}{\tau}\left(\left|n\right| - \left|m\right| x\right)
\label{eqsz}
\end{eqnarray}
Again, we can integrate $y(\tau)$ in terms of $z(\tau)$, obtaining
\begin{equation}
y(\tau)=2(\left|n\right|-\left|m\right|)\log\tau-2\log z(\tau)+k
\end{equation}
We see from this equation
that for consistency, $z(\tau) \sim \tau^{n-m}$ for $\tau \to \infty$
in contrast
with what happens for
 the
$\phi$-string. Concerning
the gauge field boundary condition, one has
 $\lim_{\tau \to \infty}x(\tau) = 1$.

As a summary of the
numerical analysis of the BPS solutions, one should note
that
 axionless string solutions found in (\cite{Hong})-(\cite{JW}) are not much modified
 by the axion which, however, contributes to the electric charge
 of the string configuration.  These explicit solutions could be of interest in the
 context of cosmic strings and, due to the coupling to the axion and  their
electric charge, their dynamics could be very different of that of
ANO vortices.

\vspace{1 cm}

\noindent Acknowledgment: I would like to thank the organizers, lecturers and
participants, and in
particular  Prof. Pedro Labra\~na, for the splendid days I spent
during the
Valparaiso school. This work was partially supported
by PIP-CONICT, PICT-ANPCYT,
UNLP and CICBA grants.


\end{document}